\documentclass[preprint,12pt]{elsarticle}
\usepackage{graphicx}
\usepackage{amssymb}
\usepackage{amsmath}
\usepackage{graphics}
\usepackage{subcaption}
\usepackage{siunitx}
\usepackage{enumerate}
\usepackage{cases}
\usepackage{textcomp}
\usepackage{xcolor}
\usepackage{booktabs}
\usepackage{lineno}
\usepackage{mathtools}
\usepackage[normalem]{ulem}
\usepackage[ruled,vlined]{algorithm2e}
\usepackage{courier}

\definecolor{darkgreen}{rgb}{0.0, 0.6, 0.0}
\newcommand{\reva}[1]{\textcolor{black}{#1}}
\newcommand{\revb}[1]{\textcolor{black}{#1}}
\newcommand{\revc}[1]{\textcolor{black}{#1}}

\newenvironment{eqsys}{\begin{equation}\begin{dcases}}{\end{dcases}\end{equation}}

\journal{Journal}
\begin{document}
\begin{frontmatter}

\title{Efficiency gains of a multi-scale integration method applied to a scale-separated model for rapidly rotating dynamos\tnoteref{mytitlenote}}
\tnotetext[mytitlenote]{This work was supported by the European Research Council (ERC) under the European Union’s Horizon 2020 research and innovation programme (grant agreement no. D5S-DLV-786780). This work was undertaken on ARC3, part of the High Performance Computing facilities at the University of Leeds, UK.}

\author[leeds]{Krasymyr Tretiak\corref{cor}}
\ead{k.tretiak@leeds.ac.uk}

\author[eth]{Meredith Plumley}
\ead{mplumley@ethz.ch}

\author[cu]{Michael Calkins}
\ead{michael.calkins@colorado.edu}

\author[leeds]{Steven Tobias}
\ead{s.m.tobias@leeds.ac.uk}

\address[leeds]{School of Mathematics, University of Leeds, United Kingdom}
\address[eth]{Institute of Geophysics, Eidgenössische Technische Hochschule Zürich, Switzerland}
\address[cu]{Department of Physics, University of Colorado, USA}
\cortext[cor]{Corresponding author}

\begin{abstract}
Numerical geodynamo simulations with parameters close to an Earth-like regime would be of great interest for understanding the dynamics of the Earth's liquid outer core and the associated geomagnetic field. 
Such simulations are far too computationally demanding owing to the large range in spatiotemporal scales.
This paper explores the application of a multi-scale timestepping method to an asymptotic model for the generation of magnetic field in the fluid outer core of the Earth. 
The method is based on the heterogeneous multiscale modelling (HMM) strategy, which incorporates scale separation and utilizes several integrating models for the fast and slow fields. 
Quantitative comparisons between the multi-scale simulations and direct solution of the asymptotic model in the limit of rapid rotation and low Ekman number are performed. 
The multi-scale method accurately captures the varying temporal and spatial dynamics of the mean magnetic field at lower computational costs compared to the direct solution of the asymptotic model.
\end{abstract}


\end{frontmatter}


\section{Introduction}
\subsection{Multiscale modeling}
Dynamical systems with various processes taking place on largely different temporal and space scales are commonly encountered in scientific and engineering problems. Such systems typically include multiple physical models which couple on different levels, either analytically or numerically, and the description of these dynamical systems usually requires a multilevel approach which is called multiscale modeling \cite{Multiscale_modelling_Brandt, WE_2011}. Multiscale modeling, therefore,  provides a method for the simultaneous modeling of connected processes at different scales and continuous information exchange between distinct models. The widespread use of multiscale modeling is driven by the need to simulate systems with high complexity that are poorly described by one-scale models and the rapid development of high-performance computing resources. 

Over the past few decades, a large number of various multiscale algorithms have been developed \cite[see][chapter 3]{WE_2011}, including methods incorporating different solution strategies which are applied to model multi-physics problems; examples of these include problems in fluids \reva{\cite{HMM_fluids,HMM_fluids_Malecha}}, gas dynamics \cite{HMM_gas_dyn}, solids \cite{HMM_solids}, molecular dynamics \cite{HMM_molecular} and others. From a wide range of different multiscale methods it is possible to distinguish two general methodologies that define common steps in construction principles and in coupling of macroscale and microscale models together: the heterogeneous multiscale method (HMM) introduced in \cite{HMM_orig_2003, HMM_orig_2005, HMM_review_2007} and the equation-free approach \cite{EqFree_kevrekidis_2003, EqFree_2009}. 
The HMM and equation-free approaches have the same overall idea based on scale separation; they utilise a macrosolver employing a large integration time step and a microsolver where the full multi-dimensional system is integrated for a short time  using a small integration time step. 
Both methods evolve the macro state of the system through the micro model and allow for information exchange between the states. 
The HMM has two main components: a macroscopic solver and an estimator, where the latter is responsible for the microscale simulations and estimation of missing data. The HMM framework solves the approximated macroscale model with the help of extracted data from the supplementary microscale model, whereas the macrostate provides the limits for building up the micro model and ensures that the microscale solution stays in an environment that is consistent with the macroscale states.
The equation-free approach addresses a way of modelling systems at a coarse level using fine-scale simulations. In certain cases, when the explicit derivation of the macro scale model is unavailable, macroscopic equations can be bypassed by using appropriately initialized short time and  fine-scale modelling. The coarse time-stepper is the key element of the equation-free approach which performs a time step of an unavailable macroscopic model.
It consists of  the following stages: lifting, simulation, restriction and extrapolation. The lifting procedure is responsible for the creation of microscale initial conditions consistent with the macrostate, whereas the restriction operation  obtains the coarse states from the fine-scale simulations.
The main difference between the heterogeneous multiscale method and equation-free methods is that the HMM starts with the macroscale model whilst the equation-free approach starting point is the microscale model. The latter uses the microscale model to retrieve the macro states. In addition, the equation-free approach has an extrapolation stage (projective integration) which computes time derivatives of the macro variables based on the microscale models and uses those calculations to extrapolate the macro variables for a large time step.

In the simplest case, a multiple time scale system with the slow dynamics $x$ evolving on a time scale of order $\mathcal{O}(1)$ and the fast dynamics $y$ with a time scale of order $\mathcal{O}(1/\varepsilon)$ can be described by the following set of equations
\begin{eqsys}\label{multiple_time_scale_system}
    \dfrac{dx}{dt} = f(x,y), \\
    \dfrac{dy}{dt} = \dfrac{1}{\varepsilon}\phi(x,y),
\end{eqsys}

\noindent where the scale separation is defined by the parameter $\varepsilon$. The small parameter $\varepsilon \ll 1$ in the system (\ref{multiple_time_scale_system}) leads to a numerically stiff problem where approximation of the slow and fast dynamics can be computationally prohibitive owing to imposed restrictions for a time step to be of $\mathcal{O}(\varepsilon)$ or smaller. Such a class of dynamical system, with several temporal scales, can be efficiently solved  by various fast integrators for stiff ODEs \cite[see][]{VSS_HMM,lee2015fast,Tao2010,FATKULLIN2004605} constructed within the framework of the heterogeneous multiscale methods. The seamless version of HMM has been introduced by E at al. \cite{E2009} with the main advantage of not requiring the reinitialization of the microscale model at each macro iteration step. Similar to HMM-like strategies, there is another class of multiscale integrators called projective integration methods \cite{PI_origin} which fit the framework of the equation-free approach \cite{EqFree_2009}. The projective integration methods were designed as explicit methods for stiff systems of ordinary differential equations and extend to higher orders based on Runge-Kutta schemes \cite{PI_2nd_order,PI_high_order}. The above mentioned method processes both the slow and fast variables at the same time without scale separation and utilizes extrapolation of all the variables for large time steps \cite{Vanden-Eijnden2007}.   

In this paper, we construct a general multi-scale method for a system of partial differential equations (PDEs) with a stiff timescale separation. The motivation for the model is introduced in the next section.
\subsection{Introduction to the Geodynamo}

In this paper, we consider the application of multi-scale timestepping methods  to an asymptotic model for the generation of magnetic field in the fluid outer core of the Earth; the so called geodynamo problem. This is a prototypical problem for testing such methods as the fluid outer core demonstrates dynamics on a vast range of timescales \citep{tobias2019}, as described below.

The rotation period $(T_\Omega)$ of the Earth is  of course one day; this is the shortest timescale of interest here. The dynamo that generates the Earth's magnetic field is believed to be driven by convection for which a typical convective time $T_c$  is on the order of $10-100$~years. In the absence of a driving source, the geomagnetic field would diffuse away on a timescale $T_\eta \sim 10^4$ years. The longest timescales of interest are given by those that describe  viscous and thermal diffusion ($T_\nu$ and $T_\kappa$ respectively); these are extremely long $T_\kappa \sim T_\nu \sim 10^9 - 10^{10}$ years. The ratio of these timescales is given by a set of  non-dimensional parameters. For the purposes of this paper, the most important of these parameters is the ratio of the rotation period to the viscous diffusion time which is known as the Ekman number, $E$, defined as 
\begin{equation}
 E = \frac{T_\Omega}{T_\nu}\sim 10^{-15}.
\label{Ekman}
\end{equation}
Another important small parameter is the magnetic Prandtl number ($Pm$) of the fluid core \revb{\citep{Roberts2013}} which is given by
\begin{equation}
 Pm = \frac{T_\eta}{T_{\nu}} \sim \revb{10^{-6}}.
\label{mag_Pm}
\end{equation}
\reva{The classical Rayleigh number is}
\begin{equation}
    \reva{Ra = \dfrac{g \gamma \Delta \hat{T} H^3}{\nu \kappa }.}
\end{equation}
\reva{Here, $g$ is the magnitude of the gravitational
acceleration, $\gamma$ is the thermal expansion coefficient, $\Delta \hat{T}$ represents the temperature difference
across the fluid layer, $H$ is a typical length scale of the system, $\nu$ is the viscosity and $\kappa$ is the thermal diffusivity of the fluid.} \revb{The reduced Rayleigh number is defined as $\widetilde{Ra} = E^{4/3} Ra$.}

The vast separation of timescales in the geodynamo problem makes computational modelling of it extremely difficult. A direct numerical simulation (DNS) approach would resolve all  temporal and spatial turbulent scales and gives the most accurate results with high resolution. However these simulations are restricted by computational resources and, therefore, simulation parameters many orders of magnitude away from those pertaining to  the Earth’s outer core must be used. In contrast to the DNS, multi-scale modelling takes advantage of the separation of scales to enable the  use of different models for each scale. Although simulations of rotating convection interacting with  magnetic fields in a spherical shell are able to produce dynamo models that bear resemblance to the observed magnetic field \citep{Glatzmaiers2_1995} there is a massive discrepancy between the parameters that can be reached in numerical models, and those that are relevant for the Earth. For instance, a recent study has reached an Ekman number as low as $E = 10^{-7}$, but the use of such small Ekman numbers requires substantially longer simulation time \citep{sjnf2017}.

In the first of these models \citep{agf2017} the parameters are selected so as to remain as close as computationally possible to the values in the Earth whilst maintaining the relevant (so-called MAC) balance. These remarkable computations were carried out at an Ekman number $E = 10^{-8}$ -- impressive, though still orders of magnitude from the correct value. \citet{sjnf2017} managed to reach $E \sim 10^{-7}$ and find an interesting inhomogeneity of the solution linked to the presence of the solid inner core combined with the effects of strong rotation. Again however it should be noted that the parameters considered are, owing to computational limitations, many orders of magnitude from the geophysically relevant regime.

In a recent series of papers a complementary approach has been developed \citep[see][and the references therein]{Calkins2015,Calkins2016a,Calkins2016b,plumley2018}. This approach involved making use of the separation of timescales in conjunction with a, less critical, separation of spatial scales to derive reduced partial differential equations that are formally valid in an asymptotic wedge in parameter space. The model, which has been developed in Cartesian geometry, involved the separation of variables into those that are rapidly varying (i.e. on a fast convective timescale) and those that are slowly varying (on either a magnetic diffusion or a viscous diffusion time). The complicated dynamics is then described by separate evolution equations for the slow and fast dynamics, which take the form of three-dimensional PDEs. In its simplest form, where a spatial horizontal planform is selected for the form of the solutions the system takes the form of a one-dimensional partial differential equation for the evolution of these fields on the relevant timescales (as described in the next section). The aim of the paper is to develop efficient solvers that are applicable to systems of this type.

The paper is organized as follows. In Section \ref{main_eq}, we briefly summarize the investigated multiscale problem: the \reva{quasi-geostrophic dynamo model (QGDM)}  main equations and previous results. In Section \ref{HMM_section}, we present our HMM-like algorithm, show its structure and main steps. Section \ref{Numerical_results} shows the results from the application of heterogeneous multiscale modelling strategy to QGDM problem and studies the performance at different Rayleigh and small Ekman numbers. Summaries of the application of our algorithm and reduction of  computational effort needed to solve the QGDM are presented in Section \ref{conclusions}.

\section{The self-consistent dynamo problem}\label{main_eq}

\subsection{The QGDM Equations}

The detailed derivation of the non-linear QGDM is given in \cite{Calkins2015}.  Here we focus on the single amplitude (SA) version of the model which was derived in \cite{plumley2018}; this model is sufficient for the purposes of demonstrating the efficacy (or otherwise) of the timestepping strategies derived here. Briefly, the multiscale model is
based on the idea of splitting dependent variables into mean  and fluctuating parts, i.e. for any variable $f$
\begin{equation}\label{key}
f = \bar{f} + f',
\end{equation}
where an overbar denotes a mean (or slowly varying) quantity and a prime denotes a (rapid) fluctuation about that mean. The model is predicated on the introduction of three temporal scales so that 
\begin{equation}\label{key}
\partial_t \rightarrow \partial_t + \epsilon^{3/2} \partial_\tau + \epsilon^{2} \partial_T,
\end{equation}
where $ t $ refers to fast convective time scale, $ \tau = \epsilon^{3/2} t $ is a slow intermediate mean magnetic field diffusion time scale\reva{,} $ T = \epsilon^{2} t $ is a slow mean temperature time scale \reva{and $\epsilon$ is a small parameter defined as}
\begin{equation}\label{ekm_eps}
    \reva{\epsilon = E^{1/3}.}
\end{equation}
The reduced Cartesian system describes the evolution of the vertical, $z$, structure and time evolution of the mean magnetic field $ \bar{\mathbf{B}}=(\bar{B}_x,\bar{B}_y) $, mean temperature $ \bar{\theta} $ and its fluctuating part $ \Theta $, vertical velocity $ W $ and the geostrophic stream function $ \Psi $ \citep{plumley2018} via 
\begin{equation}\label{eqPsi}
\partial_t\Psi + \dfrac{1}{k^2}\partial_zW = -\dfrac{\reva{\widetilde{Pm}}}{2}(\bar{B}_x^2+\bar{B}_y^2)\Psi - k^2\Psi,
\end{equation}
\begin{equation}\label{eqW}
\partial_t W + \partial_z\Psi = \dfrac{\reva{\widetilde{Ra}}}{Pr} \Theta - \dfrac{\reva{\widetilde{Pm}}}{2}(\bar{B}_x^2+\bar{B}_y^2)W - k^2W,
\end{equation}
\begin{equation}\label{eqThe}
\partial_t\Theta + W \partial_z \bar{\theta} = -\dfrac{k^2}{Pr} \Theta,
\end{equation}
\begin{equation}\label{eqtet}
\epsilon^{-2}\partial_t\bar{\theta} + \partial_z(W\Theta) = \dfrac{1}{Pr} \partial_{zz} \bar{\theta},
\end{equation}
\begin{equation}\label{eqBx}
\epsilon^{-3/2}\partial_t\bar{B}_x = -\reva{\widetilde{Pm}} \; \partial_z(\Psi W \bar{B}_y) + \dfrac{1}{\reva{\widetilde{Pm}}} \partial_{zz}\bar{B}_x,
\end{equation}
\begin{equation}\label{eqBy}
\epsilon^{-3/2}\partial_t\bar{B}_y = \reva{\widetilde{Pm}} \; \partial_z(\Psi W \bar{B}_x) + \dfrac{1}{\reva{\widetilde{Pm}}} \partial_{zz}\bar{B}_y.
\end{equation}
\reva{In this set of equations, all the parameters including} the Prandtl number $Pr$, \reva{the  reduced  Rayleigh  number} and the \reva{reduced} magnetic Prandtl number \reva{$\widetilde{Pm} = \epsilon^{-1/2} Pm$} are $\mathcal{O}(1)$.

We impose the following boundary conditions for the mean magnetic field
\begin{equation}\label{key}
\bar{B}_x=  \bar{B}_y= 0 \quad \text{at} \quad z = 0,1,
\end{equation}
vertical velocity
\begin{equation}\label{key}
W = 0 \quad \text{at} \quad z = 0,1,
\end{equation}
and mean temperature
\begin{equation}\label{key}
\bar{\theta} = 1,0  \quad \text{at} \quad z = 0,1.
\end{equation}
Initial conditions are set as
\begin{equation}\label{key}
 \Psi(0,z) = -\pi/k^4 \cos(\pi z), \quad W(0,z) = \sin(\pi z), \quad \Theta(0,z) = 1/k^2 \sin(\pi z),
\end{equation}
\begin{equation}\label{key}
 \bar{\theta}(0,z) = 1 -z, \quad \bar{B}_x(0,z) = \sin(\pi z), \quad \bar{B}_y(0,z) = \sin(\pi z).
\end{equation}
In addition, we implement the steady-state approximation  for the mean temperature $ \epsilon^{-2}\partial_t\bar{\theta} = 0 $ employed in \cite{plumley2018} and simplify  (\ref{eqtet}) to the diagnostic equation
\begin{equation}\label{eqtet2}
\partial_z(W\Theta) = \dfrac{1}{Pr} \partial_{zz} \bar{\theta}.
\end{equation}
\reva{In the system (\ref{eqPsi}-\ref{eqBy}) the mean temperature $ \bar{\theta}$ evolves on much longer timescale than the magnetic field therefore we can assume that $ \bar{\theta}$ is constant and use equation (\ref{eqtet2}).}
 
In order to assess and analyse the numerical solutions of the QGDM system we check several important quantities. The Nusselt number $ Nu $ is the ratio of convective to conductive heat transfer through the boundary fluid layer
\begin{equation}\label{nu_value}
Nu = 1 + Pr\langle W \Theta \rangle,
\end{equation}
where angle brackets denote vertical averaging
\begin{equation}
\langle f \rangle =  \dfrac{1}{L} \int_{0}^{L} f(z)dz.
\end{equation}{}
Another useful diagnostic is the non-dimensional mean magnetic energy $ E_M $, defined as
\begin{equation}\label{me_value}
E_M = \dfrac{1}{2} \langle \vec{B} \cdot \vec{B} \rangle.
\end{equation}

\subsection{The QGDM Dynamics}

\reva{The dynamics we investigate here are quasilinear.  The problem is of the second type as considered by Michel $\&$ Chini \cite{Michel2019}, i.e. it is a quasilinear system in which the fluctuations would grow exponentially fast in the absence of feedback on the slow variables. In this case the dynamics takes the form of rapid growth when a marginal state is not achieved through adjustment of the slow variable. The fast dynamics also has a slow component because of the slow changes in the mean leading to changes in the growth-rate of the quasilinear fast dynamics. This type of behaviour makes the separation of slow and fast dynamics much more difficult than say for the case of fast waves riding on a slowly varying background. Indeed the fast dynamics for an out-of-equilibrium slow variable can really only be evaluated at the start of the calculation. 
In this limited period of the evolution one can begin to see the asymptotic behaviour as $\epsilon$ gets small.}

\begin{figure}[t!]
    \centering
	\begin{subfigure}{0.45\textwidth}
		\includegraphics[width=\linewidth]{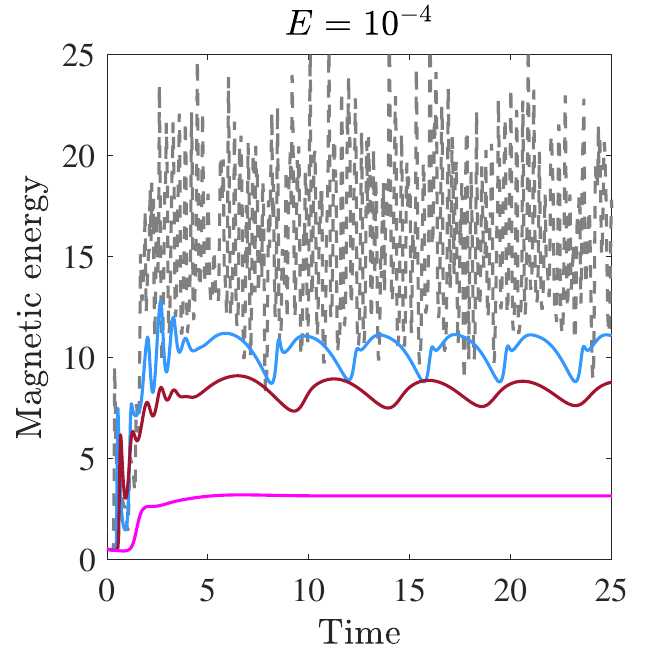}
		\caption{}
	\end{subfigure}
	\begin{subfigure}{0.45\textwidth}
		\includegraphics[width=\linewidth]{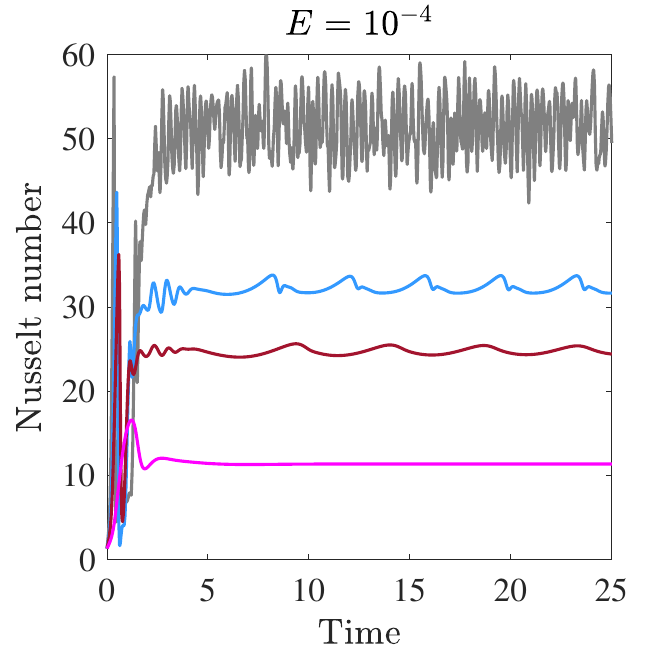}
		\caption{}
	\end{subfigure}
	\begin{subfigure}{0.45\textwidth}
		\includegraphics[width=\linewidth]{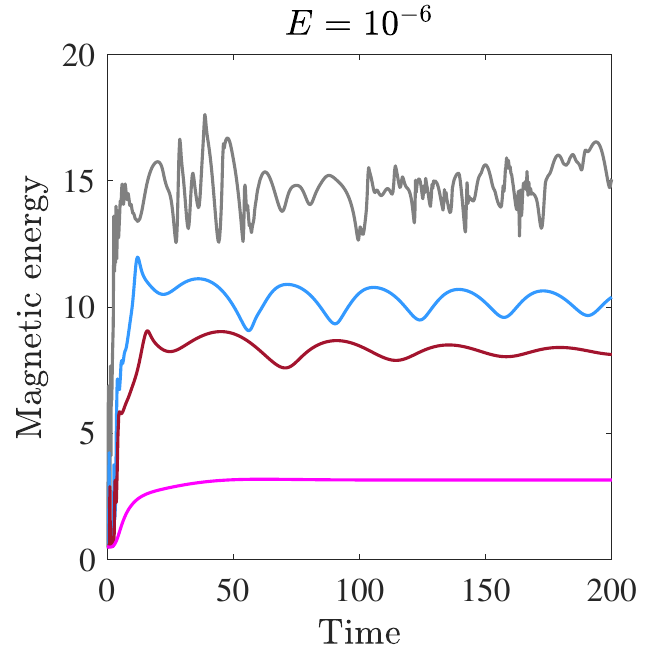}
		\caption{}
	\end{subfigure}
	\begin{subfigure}{0.45\textwidth}
		\includegraphics[width=\linewidth]{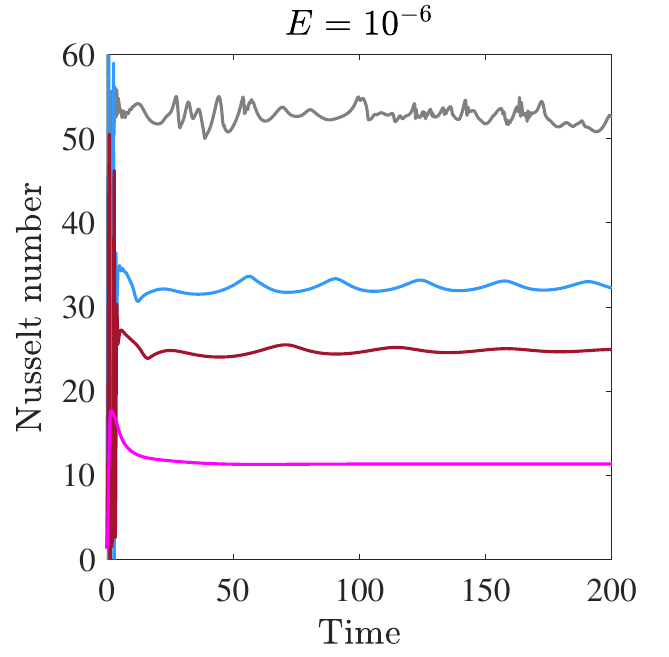}
		\caption{}
	\end{subfigure}
	\caption{\reva{Comparison of the magnetic energy (left) and the Nusselt number (right) for different values of $\widetilde{Ra}$. The simulation parameters are $Pr=1$, $\widetilde{Pm}=0.7$, $(a,b)$ $E=10^{-4}$, $(c,d)$ $E=10^{-6}$. $\widetilde{Ra}=40$ -- magenta,  $\widetilde{Ra}=80$ -- dark red, $\widetilde{Ra}=100$ -- light blue and $\widetilde{Ra}=150$ -- grey.}}
	\label{fig:main_param}
\end{figure}

Fig. \ref{fig:main_param} shows the dynamics of the QGDM via timeseries for the magnetic energy and the Nusselt number for four different values of \revb{reduced} Rayleigh number ($\revb{\widetilde{Ra}}$) and two choices of the Ekman number ($E$). These reference solutions were obtained by solving the stiff equations with a small timestep determined by the fast dynamics. As a result of their numerical stiffness, the solutions were expensive to calculate, with the computational expense increasing as the Ekman number is reduced.
The dynamics for both choices of Ekman number are similar, though the timescale of variability is of course different. As expected, as $\revb{\widetilde{Ra}}$ is increased past a critical value dynamo action occurs in which kinetic energy is continuously converted to magnetic energy. For low values of $\revb{\widetilde{Ra}}$ the numerical solution reaches a steady state, as shown by the behavior of the magnetic energy and the Nusselt number. A further increase in $\revb{\widetilde{Ra}}$ eventually yields a Hopf bifurcation to an oscillatory mode that itself becomes unstable to chaotically modulated solutions. Whilst the particular dynamics in this chaotic regime is sensitive to the initial conditions and accuracy of the numerical scheme, the broad level of magnetic field saturation and heat transport is not. It is these broad dynamics that we attempt to replicate with our faster multi-scale methods.

\section{Time stepping strategy}\label{HMM_section}
\subsection{The heterogeneous multiscale method}
In this section we present the HMM-like strategy for solving the simplified multiscale QGDM problem.  The key idea is to develop a numerical approximation for the slow evolution of the original system, but with less computational effort than the stiff system. Our approximation scheme is constructed on the basis of available time-stepping methods that are accurate up to 3rd order with respect to small time step inside the open source DEDALUS framework \cite{dedalus} and we utilize flexible built-in tools to analyse the results. \reva{Here, the  HMM-like  strategy is similar to the one used by Malecha~\textit{et~al.}~\cite{HMM_fluids_Malecha} for solving coupled systems of PDEs.}  

For simplicity, we denote the investigated system (\ref{eqPsi}-\ref{eqBy}) (using the simplification stated by equation (\ref{eqtet2})) as
\begin{subnumcases}{}
\dot{F} = f(F,S) \label{ODE_fast}\\
\dot{S} = \varepsilon g(F,S) \label{ODE_slow}
\end{subnumcases}
where the equation for $ F $ represents the fluctuating temperature and velocity, $ S $ refers to the slow dynamics including the equation for the mean magnetic field, $ \varepsilon = \epsilon^{3/2} $ is the small parameter which defines scale separation \reva{and $\epsilon$ is defined in formula (\ref{ekm_eps})}. A basic algorithm for solving \reva{this} system (\ref{ODE_fast}-\ref{ODE_slow}) with a reduced time stepping strategy is presented in Fig. \ref{fig:HMMscheme} and can be summarized by the following four-step procedure:
\begin{figure}[t]
    \centering
	\begin{subfigure}{0.6\textwidth}
	\includegraphics[width=\linewidth]{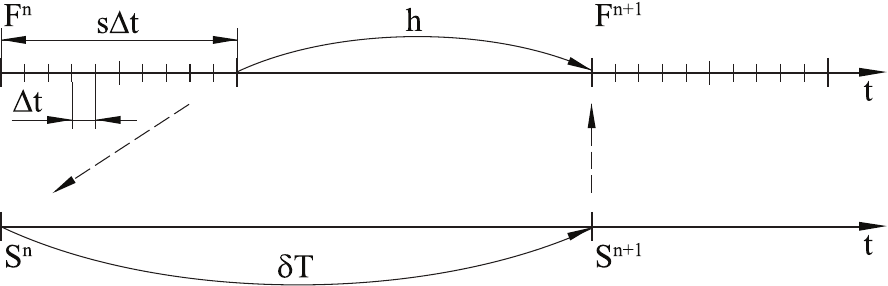}
	\end{subfigure}
	\caption{\reva{A schematic depicting} the time steps taken by the micro and macro solvers in the interval $[t_n,t_{n+1}]$ where $S,F$ are the slow and fast states, $\Delta t$ is the micro step, $s$ defines number of micro steps, $\delta T$ is the macro step, $h$ is projecting step which updates fast states to $t_{n+1}$.}
	\label{fig:HMMscheme}
\end{figure}

\begin{enumerate}
	\item \textit{micro-solver}: with given $ S = S^n $, $ F = F^n $ integrate  the equation (\ref{ODE_fast}) for the fast variable by taking \reva{$s$ time steps of length} $ \Delta t $ and keeping fixed the slow variable $ S^n $
	
	\item \textit{estimator}: observe the dynamics of the fast variable and estimate it's rate of change by getting the time averaged force $ \widetilde{F} $.  The averaging technique can be performed as a simple mean
	\begin{equation}\label{simple_mean}
    \reva{\widetilde{F}} = \dfrac{1}{s\Delta t}\int_{t_n}^{t_n + s\Delta t}F(\tau)d\tau
	\end{equation}
	or using a kernel with some special properties
	\begin{equation}\label{KDE}
	\reva{\widetilde{F}_K} = \dfrac{\int_{t_n}^{t_n + s\Delta t} F(\tau) K(\tau)d\tau}{\int_{t_n}^{t_n + s\Delta t} K(\tau)d\tau}.
	\end{equation}
	Formula (\ref{KDE}) computes a weighted average for all data points that are within the window $[t_n, t_n + s\Delta t]$ and the function $ K(u)$ defines how the weights change with the distance inside $[t_n, t_n + s\Delta t]$. Several examples of how the choice of averaging technique affects the accuracy of the methods are shown in Table \ref{table:kernels}-\ref{table:kernels2} and discussed in Section \ref{Numerical_results}.

	\item \textit{macro-solver}: with given $ S = S^n $, and averaged $ F = \tilde{F} $, integrate the slow equation (\ref{ODE_slow}) by taking one large step 
	\begin{equation}\label{slow_step}
	\delta T  = f s \Delta t,
	\end{equation}
	where $ f \geq 1 $ is scaling factor. This parameter defines how large the slow step will be compared to the averaging window. The case $ f = 1 $ leads to the scheme which was used in \cite{plumley2018}.
	
    \item \textit{projector}: with given $ S = \reva{S^{n+1}} $ update the solution for the fast states at $ t = t_{n+1} $ by taking one step
    \begin{equation}
    h = \delta T - s\Delta t = (f-1)\, s\Delta t.
    \end{equation}
    \reva{Here we exploit the idea of projective integration methods for systems with multiple time scales \cite{PI_origin, Vanden-Eijnden2007}.}
\end{enumerate}
\reva{Projective integration method observes the dynamics of variables in order to estimate their rate of change and computes a chord slope. Then it uses these estimates to extrapolate evolution of variables over a larger time step. In our scheme projection step is similar to the extrapolation stage in projective integration method. Relatively large step $h$ is possible since in the system (\ref{eqPsi}-\ref{eqBy}) the fast dynamics has a slow component which is imprinted by the slow variables onto the quasilinear terms in equations for the fast variables.}

\begin{figure}[ht!]
	\begin{subfigure}{0.49\textwidth}
		\includegraphics[width=\linewidth]{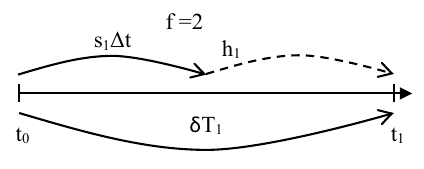}
		\caption{}
		\label{fig:scheme_two_a}
	\end{subfigure}
	\begin{subfigure}{0.49\textwidth}
		\includegraphics[width=\linewidth]{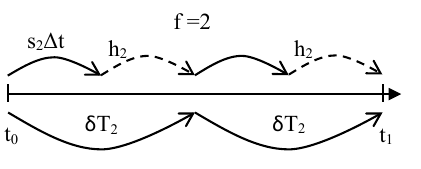}
		\caption{}
		\label{fig:scheme_two_b}
	\end{subfigure}
	\caption{A schematic representation of the integration scheme for fixed $\Delta t$, $f=2$ and $s_1 = 2s_2$ where $\Delta t$ is micro step, $s$ defines number of micro steps, $\delta T$ is macro step and $h$ is the projecting step which updates fast states to $t_{1}$.}
	\label{fig:scheme_two}
\end{figure}

Equation (\ref{slow_step}) shows the functional dependence $\delta T$ on the two variables $f$, $s$ for a fixed $\Delta t$. A schematic of the visual change of the slow step $\delta T$ is shown in Fig. \ref{fig:scheme_two}, where we apply our scheme to integrate the equations on an interval $[t_0,t_1]$ with scaling factor $f=2$ which means that $\delta T$ is two times larger than the size of averaging window on the fast scale. Parameters $s_1$, $s_2$ are defined such that $s_1/s_2 = 2$ and Fig. \ref{fig:scheme_two_a}, \ref{fig:scheme_two_b} illustrate how the choice of $s$, indicating the number of steps, will affect the overall scheme for fixed $f$. Decreasing the number of the fast steps $s$ will decrease not only the size of averaging window but also $\delta T$. Scheme $(b)$ on Fig. \ref{fig:scheme_two} with the smaller $s_2$ requires nearly the same computational effort to integrate the system on interval $[t_0,t_1]$ by making two global steps, but will produce more accurate results due to smaller steps $h_2$ when updating the fast variable states. 

The procedure described above includes linearisation of the problem when averaging the fast convective steps. We assume that the slow field $ S $ is changing slowly over the course of fast time stepping and the nonlinearity is therefore weak. Such an approach is a relatively fast way of getting an approximate solution of the slowly varying magnetic field. The choice of scaling factor $f$ strongly depends on the parameters of the system as well as on the time stepping technique for both the macro and micro solvers. It has been determined empirically  that the possible values of $f$ are within the interval $[1,4]$. Larger values produce numerically unstable results.

\revc{Solving systems of PDEs in DEDALUS requires creating an \texttt{Problem} object which represents the problem type, domain, equations, parameters, initial and boundary conditions, and a temporal integration scheme. We use three predefined objects to solve the scale-separated problem for rapidly rotating dynamo.}
\revb{We exploit the 1\textit{st} \texttt{Problem} object (\textit{micro-solver}) to get the solution of equations (\ref{eqPsi}-\ref{eqtet}) on the fast scale and the 2\textit{nd} object (\textit{macro-solver}) solves the equations (\ref{eqBx}-\ref{eqBy}) for the mean magnetic field.} \revb{DEDALUS framework has the base class for implicit-explicit (IMEX) multistep methods that supports the globally stiffly accurate family of IMEX Runge-Kutta methods.} Both solvers utilize 3\textit{rd} order 4\textit{th} stage implicit-explicit Runge-Kutta scheme \textit{RK443} \cite{rk443}. The 3\textit{rd} object replicates the equations (\ref{eqPsi}-\ref{eqtet}) and updates the fast states to time $ t_{n+1}$ by making step $h$ after the slow step being made by the \textit{macro-solver}. Here we apply 1\textit{st}-order Euler extrapolation and use semi-implicit BDF scheme \cite{sbdf2}. 
\revc{The problem is solved by grouping solvers into main loop and iteratively applying the temporal integration scheme for each solver assuming data interaction as integration progresses.} The temporal integration scheme for the \textit{micro}- , \textit{macro}- solvers can be selected independently as well as semi-implicit BDF scheme for the projection step, however such combination shows optimal stability and accuracy properties comparing to other available options. The scaling factor $ f $  and number of steps $ s $ on the fast scale strongly depend on the model parameters $ \revb{\widetilde{Ra}}, E, k$. In the results section \ref{Numerical_results}, we show how the scaling factor $ f $ and number of micro steps $ s $ (or size of the averaging window $ s \Delta t $) affects the accuracy of the solution, total runtime and overall performance. 
\section{Numerical results} \label{Numerical_results}

In this section, we apply our scheme to the system (\ref{eqPsi}-\ref{eqBy}) with the simplification (\ref{eqtet2}). We provide numerical results for different values of $E$, $ \revb{\widetilde{Ra}} $ and compare them with the reference solutions. The reference solutions were obtained by solving the same equations (\ref{eqPsi}-\ref{eqBy}) \reva{and assuming the steady state approximation (\ref{eqtet2})} as one system using Runge-Kutta method and processing both slow and fast variables at the same time with $\delta t$ step which corresponds to micro step $\Delta t$ \reva{in} Fig.~\ref{fig:HMMscheme}.

Simulations are carried out in a 1D domain with size $L = 1.0$. The vertical spatial dimension is discretized using Chebyshev polynomials and we keep constant number of Chebyshev points in all simulations $ N_z=128 $. Additional parameters of the system are set to be $ Pr=1.0$, $\reva{\widetilde{Pm}}=0.7$ and wavenumber $ k = 1.3048 $ which corresponds to steady convection \cite{chandrasekhar1961}. An appropriate choice of the smoothing weight function in the \textit{estimator} step can  positively influence the \textit{macro-solver} step in terms of stability, based on the broad set of test runs completed. Table \ref{table:kernels} and \reva{Table \ref{table:kernels2} show} accuracy comparisons of different smoothing functions with respect to simple averaging (\ref{simple_mean}).
\begin{table}[ht!]
    \centering
    \begin{tabular}{@{\extracolsep{6pt}} lcccccc } 
    \hline
    Type    & $E_{rel}$ & $\sigma_X$ & $D_{max}$ & $E_{rel}$ & $\sigma_X$ & $D_{max}$ \\
    \hline
    {} & \multicolumn{3}{c}{$\widetilde{Ra}=80$}  & \multicolumn{3}{c}{$\widetilde{Ra}=150$}  \\
    \cmidrule{2-4} \cmidrule{5-7}
     Simple mean &  0.000769  &  0.006442 & 0.0124 &  0.179 &  2.62  &  6.30\\
       Parabolic &  0.000755  &  0.006325 & 0.0121 &  0.112  &  1.64  & 4.73 \\
        Gaussian &  0.000764  &  0.006401 & 0.0123 &  0.185  &  2.70  & 6.44 \\
         Quartic &  0.000750  &  0.006278 & 0.0119 &  0.107  &  1.57  & 4.72 \\
      \reva{Triangular} &  0.000752  &  0.006298  & 0.012 & 0.114   &  1.67  &  4.72\\
    \hline
    \end{tabular}
    \caption{\reva{Accuracy comparisons of the magnetic energy for different averaging functions using  simulation parameters $E=10^{-6}$, $s=100$, $f=2.0$, $\Delta t = 5 \cdot 10^{-4}$ and $t_{sim} = 150$. Left sub-column shows errors for $\widetilde{Ra}=80$ and right for $\widetilde{Ra}=150$.}}
    \label{table:kernels}
\end{table}
\begin{table}[ht!]
    \centering
    \begin{tabular}{@{\extracolsep{6pt}} lcccccc } 
    \hline
    Type    & $E_{rel}$ & $\sigma_X$ & $D_{max}$ & $E_{rel}$ & $\sigma_X$ & $D_{max}$ \\
    \hline
    {} & \multicolumn{3}{c}{$\widetilde{Ra}=80$}  & \multicolumn{3}{c}{$\widetilde{Ra}=150$}  \\
    \cmidrule{2-4} \cmidrule{5-7}
     Simple mean &  0.00269  &  0.00793 & 0.0150 &  0.308 &  1.18  & 2.73 \\
       Parabolic &  0.00264  &  0.00781 & 0.0148 &  0.275 &  1.06  & 2.63 \\
        Gaussian &  0.00267  & 0.00789  & 0.0149 &  0.357 &  1.37  & 3.29 \\
         Quartic &  0.00263  & 0.00776  & 0.0147 &  0.324 &  1.24 &  2.89\\
      \reva{Triangular}&  0.00264  & 0.00778 & 0.0147 &  0.316  &  1.21  & 2.84 \\
    \hline
    \end{tabular}
    \caption{\reva{Accuracy comparisons of $\langle B_x \rangle$ component for different averaging functions using  simulation parameters $E=10^{-6}$, $s=100$, $f=2.0$, $\Delta t = 5 \cdot 10^{-4}$ and $t_{sim} = 150$. Left sub-column shows errors for $\widetilde{Ra}=80$ and right for $\widetilde{Ra}=150$.}}
    \label{table:kernels2}
\end{table}
Results are presented for the mean magnetic energy $E_M$ and \reva{$\langle B_x \rangle$ component defined as} 
\begin{equation}
    \reva{\langle B_x \rangle = \sqrt{\int_0^1 B_x^2 \; dz}}.
\end{equation}
The maximum deviation $D_{max}$ in $E_M$ and $\langle B_x \rangle$ at all points of the computed profile reads
\begin{equation}\label{D_max}
D_{max} := \max_{n = 0, \ldots, N} \left| x_n - x^{(\text{ref})}(t_n) \right|,
\end{equation}
with $t_n$, $n=0, \ldots, N$ being the time steps for the current resolution and $x^{(ref)}(t_n)$ the reference solution at those points. Here we use high accuracy intermediate  interpolation of the reference solution in order to obtain the  values at desired time points. We compute the relative $ l_2 $ error 
\begin{equation}\label{E_rel}
E_{rel} = \left. \left( \sum_{i=1}^{N}\left| x_i - x^{(\text{ref})}(t_i)\right|^2 \right)^{1/2} \middle/ \left\lVert x^{(\text{ref})} \right\rVert_2 \right.,
\end{equation}
where $ N $ is number of slow time steps on the interval $ [0, t_{end}] $, $ x^{(\text{ref})} $, the reference solution obtained using $ \Delta t $, and  $ x_i $, the solution with the application of reduction strategy. Standard deviation $\sigma_X$ defined as follows
\begin{equation}
    \sigma_X = \sqrt{\dfrac{1}{N} \sum_{i=1}^{N} (x_i - x^{(\text{ref})})^2}.
\end{equation}
We use the following definitions for the averaging functions: parabolic -- $K(u) =3(1-u^2)/4$ for $\lvert u \rvert \leq 1$, Gaussian -- $K(u) = \exp{(- u^2/2)}/\sqrt{2\pi}$, quartic -- $K(u) = 15(1-u^2)^2/16$ for $\lvert u \rvert \leq 1$ and triangular -- $K(u) = (1 - \lvert u \rvert)$ for $\lvert u \rvert \leq 1$ where the interval $u=[-1,1]$ corresponds to $[t_n, t_n + s\Delta t]$ in formula (\ref{KDE}). All averaging functions show relatively the same level of errors with a slight advantage in favour of the parabolic, the kernel having the smallest relative error $E_{rel}$, standard deviation $\sigma_X $ and lowest maximum defect $D_{max}$ for $\revb{\widetilde{Ra}} = 150$. \reva{For sufficiently large $\widetilde{Ra} > 100 $ solution of the system (\ref{eqPsi}-\ref{eqBy}) yields chaotic dynamo (see Fig. \ref{fig:main_param}). In this regime slow dynamics of the dynamo can not be separated from the fast dynamics and the algorithm struggles to get the right numerical approximation for the slow evolution which results in significant errors in $E_M$ and $\langle B_x \rangle$ (see Table \ref{table:kernels}-\ref{table:kernels2}).} 

\subsection{Work precision}
\begin{figure}[ht!]
\centering
	\begin{subfigure}{0.4\textwidth}
		\includegraphics[width=\linewidth]{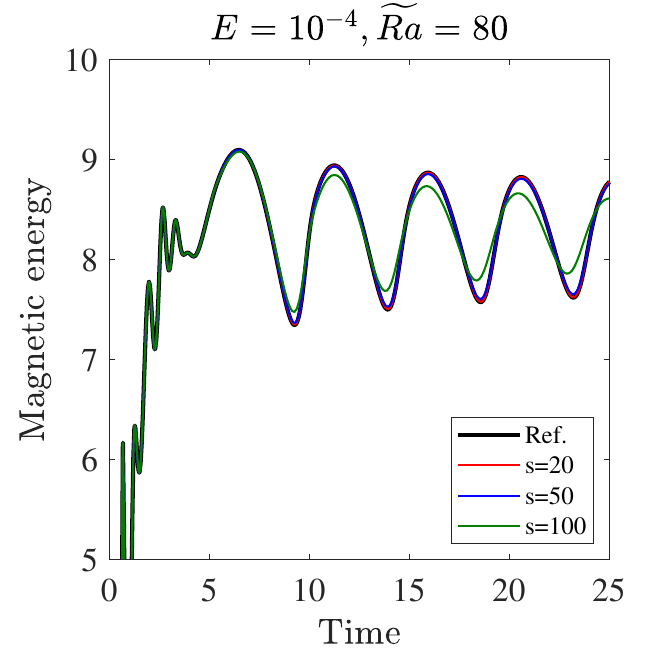}
		\caption{}
		\label{fig:ME_NU_general_a}
	\end{subfigure}
	\begin{subfigure}{0.4\textwidth}
		\includegraphics[width=\linewidth]{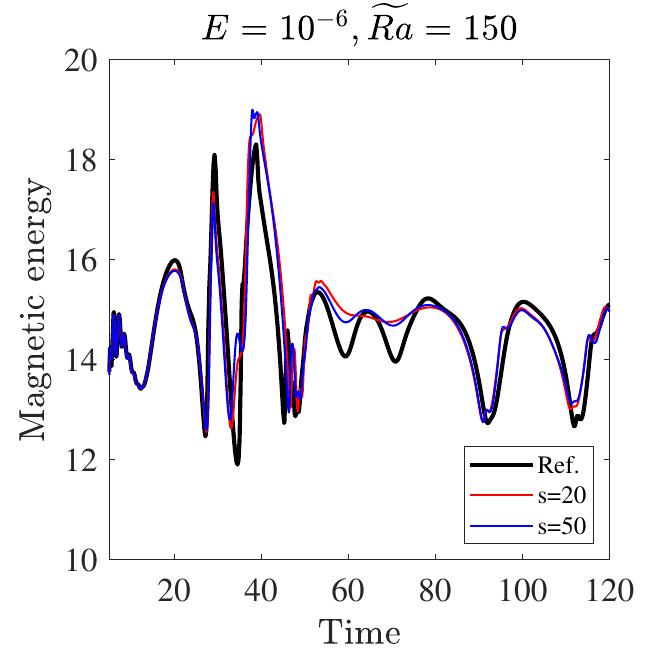}
		\caption{}
		\label{fig:ME_NU_general_b}
	\end{subfigure}
		\begin{subfigure}{0.4\textwidth}
		\includegraphics[width=\linewidth]{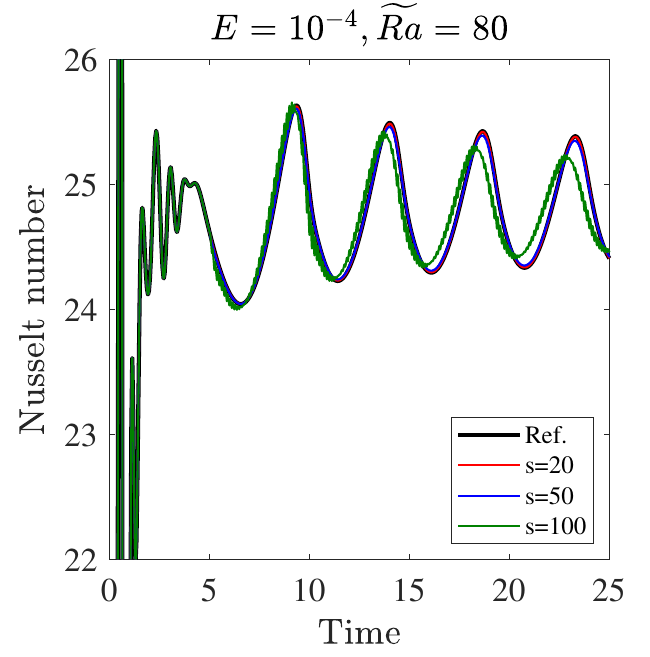}
		\caption{}
		\label{fig:ME_NU_general_c}
	\end{subfigure}
	\begin{subfigure}{0.4\textwidth}
		\includegraphics[width=\linewidth]{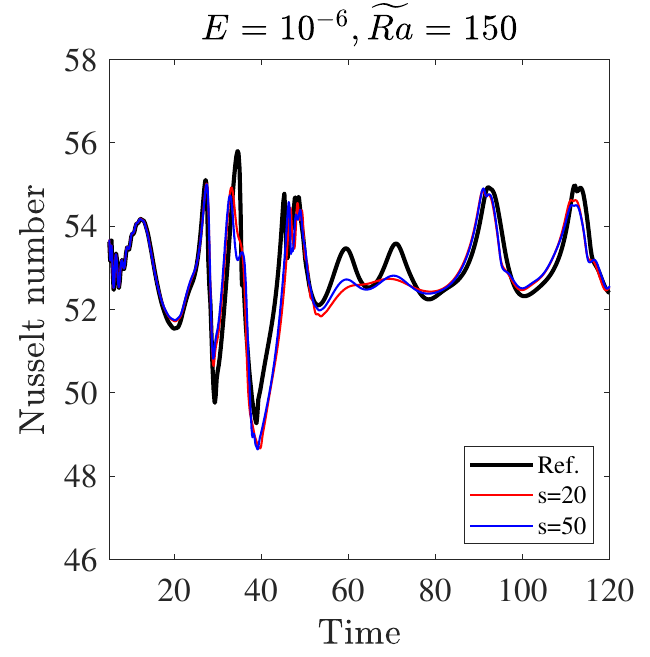}
		\caption{}
		\label{fig:ME_NU_general_d}
	\end{subfigure}
	\caption{\reva{Time evolution of the magnetic energy and the Nusselt number: $\widetilde{Ra}=80, E=10^{-4}$, scaling factor $f = 2.5$ (on the left) and $\widetilde{Ra}=150, E=10^{-6}$, scaling factor $f = 2.0$ (on the right).}}
	\label{fig:ME_NU_general}
\end{figure}

Fig.~\ref{fig:ME_NU_general} presents calculated time evolution of the Nusselt number and the magnetic energy depending on different values of $ \revb{\widetilde{Ra}} $ and $ E $ with fixed micro step $ \Delta t = 5 \cdot 10^{-4} $ for all cases. The reference solutions for both quantities are shown as the black lines. We also fixed scaling factor $f=2.5$ for $(a)$ and $f=2.0$ for $(b)$ figures which allow us to demonstrate only the influence $s$ on the accuracy of the solutions.

Figure \ref{fig:ME_NU_general_a} and  \ref{fig:ME_NU_general_c} present the results with $\revb{\widetilde{Ra}}=80, E=10^{-4}$ showing the magnetic energy on the top and  $ Nu $ at the bottom. For $ \revb{\widetilde{Ra}} \lesssim 80 $ the designed integration scheme shows good consistency with the reference solution using different sizes of the averaging window $ s \Delta t $ and constant scaling factor $ f = 2.5 $ for the slow step. For a higher number of fast steps $ s = 100 $ the results are slightly out of phase but still in good agreement with the reference. This is a consequence of the fact that the increase of $s$ extends the $\delta T$ step. For instance, the scheme with $s_1 = 20$ uses $\delta T_1 = 0.025$, $h_1 = 0.015$ and the scheme with $s_2 = 100$ uses $\delta T_2 = 0.125$, $h_2 = 0.075$, so the five times larger $s_2$ results in a five times larger $\delta T_2$. The slowly varying magnetic field allows for relatively large steps to be taken, however the stability and accuracy of the solution also depends on the length of the projector step $h$ which updates the fast states at the end of each global step $[t_n,t_{n+1}]$ \reva{in} Fig.~\ref{fig:HMMscheme}.
The method evolves the macro state of the system through the micro model and there is exchange between the states. Therefore any inaccuracies in timestepping the fast scale will result in errors on the slow scale too.  Lower Ekman numbers allow for both larger averaging windows $s$ and larger $h$ steps to be taken and still retain accurate solutions.

Figure  \ref{fig:ME_NU_general_b} and \ref{fig:ME_NU_general_d}  present time evolution profiles for the case with $\revb{\widetilde{Ra}}=150, E=10^{-6}$. As shown in Fig.\ref{fig:main_param}, higher values of $ \revb{\widetilde{Ra}} $  generate additional frequencies in both the convection and mean magnetic field scales. Sufficiently large $ \revb{\widetilde{Ra}} $ leads to a loss of periodicity and chaotic convection. A relatively small averaging window $ s = 20 $ produces stable results while integrating the system with $ \revb{\widetilde{Ra}} = 150 $. However, the simulation with $ s \geq 50 $ cannot deliver good accuracy and generates numerical instabilities in the solution which are caused by an inability to capture the right dynamics when taking large steps on the mean magnetic field and updating the fast variables during the \textit{projector's} step.

Table~\ref{table:Nudev} shows maximum deviation for the Nusselt number during simulations with fixed $ E = 10^{-6} $ and varying $ \revb{\widetilde{Ra}} $. The time-averaged reference $ Nu $ is also indicated in the table for each case.  The maximum deviation $d_{Nu}$ in $ Nu $ at all points of the computed profile calculated using formula (\ref{D_max}).
\begin{table}[ht!]
	\centering
	\begin{tabular}{@{\extracolsep{6pt}}lcccccc}
		\toprule   
		{} & \multicolumn{2}{c}{$ \revb{\widetilde{Ra}}=80 $}  & \multicolumn{2}{c}{$ \revb{\widetilde{Ra}}=130 $} & \multicolumn{2}{c}{$ \revb{\widetilde{Ra}}=150 $} \\
		{} & \multicolumn{2}{c}{(aver. $ Nu=24 $)}  & \multicolumn{2}{c}{(aver. $ Nu=44 $)} & \multicolumn{2}{c}{(aver. $ Nu=52 $)} \\
		\cmidrule{2-7}
		&  $ f=2.0 $ & $ f=3.0 $ & $ f=2.0 $ & $ f=3.0 $ & $ f=2.0 $ & $ f=3.0 $\\ 
		\cmidrule{2-3} \cmidrule{4-5} \cmidrule{6-7} 
		$ s=20 $ & 0.00821 & 0.01372 & 0.08673 & 5.21771 & 4.84354 & 16.7028\\ 
		$ s=50 $ & 0.02216 & 6.66026 & 0.20040 & 0.51769 & 6.05991 & 24.5380\\ 
		$ s=100$ & 0.04876 & 0.19758 & 0.51646 & 45.3400 & 12.1876 & 21.8959\\ 
		$ s=200$ & 0.10310 & 9.42583 & 1.07074 & 23.5082 & --      & --     \\ 
		\bottomrule
	\end{tabular}
	\caption{Maximum deviation for the Nusselt number for different $\widetilde{Ra}$, $s$ and $f$ parameters at $ E = 10^{-6} $,  $ \Delta t = 5\cdot10^{-4} $.}
	\label{table:Nudev} 
\end{table}
Table \ref{table:NuAver} shows the time-averaged Nusselt number for the above mentioned parameters.  The two measurements for error in the Nusselt number in tables ~\ref{table:Nudev} and \ref{table:NuAver} provide both the instantaneous and time-averaged accuracy.
\begin{table}[ht!]
	\centering
	\begin{tabular}{@{\extracolsep{6pt}}lcccccc}
		\toprule   
		{} & \multicolumn{2}{c}{$ \revb{\widetilde{Ra}}=80 $}  & \multicolumn{2}{c}{$ \revb{\widetilde{Ra}}=130 $} & \multicolumn{2}{c}{$ \revb{\widetilde{Ra}}=150 $} \\
		{} & \multicolumn{2}{c}{$ Nu=24.74 $}  & \multicolumn{2}{c}{ $ Nu=44.11 $} & \multicolumn{2}{c}{$ Nu=52.79 $} \\
		\cmidrule{2-7}
		&  $ f=2.0 $ & $ f=3.0 $ & $ f=2.0 $ & $ f=3.0 $ & $ f=2.0 $ & $ f=3.0 $\\ 
		\cmidrule{2-3} \cmidrule{4-5} \cmidrule{6-7} 
		$ s=20 $ & 24.71 & 24.68 & 44.59 & 41.24 & 52.29 & 41.63\\ 
		$ s=50 $ & 24.71 & 19.46 & 44.59 & 44.60 & 51.83 & 45.74\\  
		$ s=100$ & 24.71 & 24.68 & 44.60 & 21.50 & 52.06 & 50.88\\  
		$ s=200$ & 24.71 & 23.39 & 44.64 & 42.70 &  -- & -- \\  
		\bottomrule
	\end{tabular}
	\caption{The averaged Nusselt number for different $\revb{\widetilde{Ra}}$, $s$ and $f$ parameters at $ E = 10^{-6} $,  $ \Delta t = 5\cdot10^{-4} $. Values from the reference solution are indicated on the top.}
	\label{table:NuAver} 
\end{table}
For low $ \revb{\widetilde{Ra}} $ and scaling factor $ f = 2.0 $ the scheme produces highly accurate results even for $ s = 200 $ with the maximum deviation less then $ 0.5\% $. Simulations with $ \revb{\widetilde{Ra}} = 130 $ are still in good agreement with the reference for doubled step on the slow scale with maximum deviation $ \sim 2\% $. The case for $ \revb{\widetilde{Ra}} = 150 $ requires the decrease of micro step $ \Delta t $ in order to produce more accurate results. 

\begin{figure}[ht!]
	\centering
	\begin{subfigure}{0.325\textwidth}
		\includegraphics[width=\linewidth]{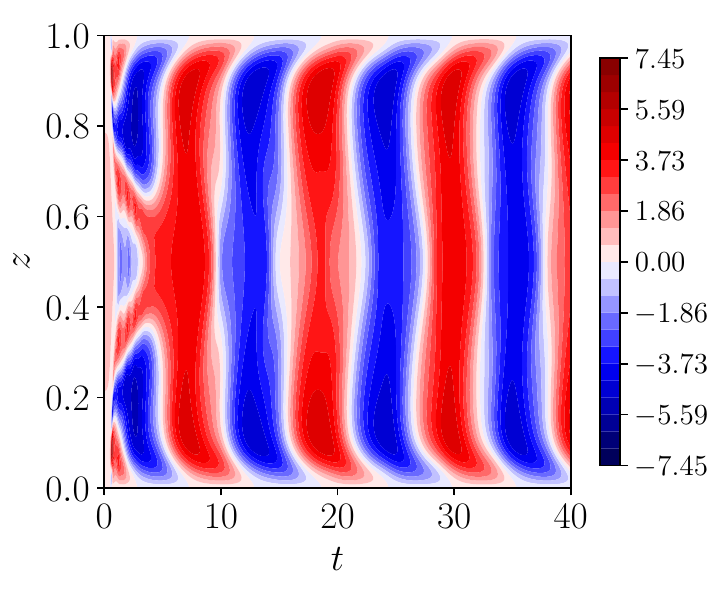}
		\caption{}
		\label{fig:contour_ekm4_ra80}
	\end{subfigure}
	\begin{subfigure}{0.325\textwidth}
		\includegraphics[width=\linewidth]{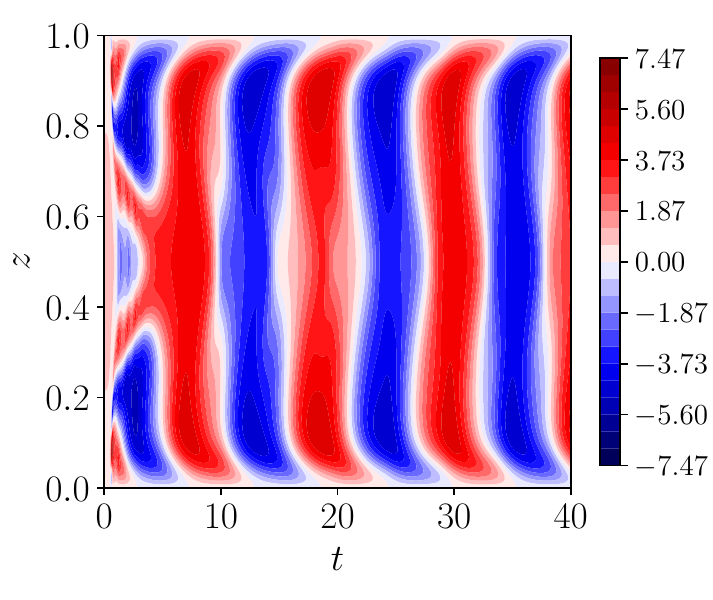}
		\caption{}
		\label{fig:contour_Ra80_s20}
	\end{subfigure}
	\begin{subfigure}{0.325\textwidth}
		\includegraphics[width=\linewidth]{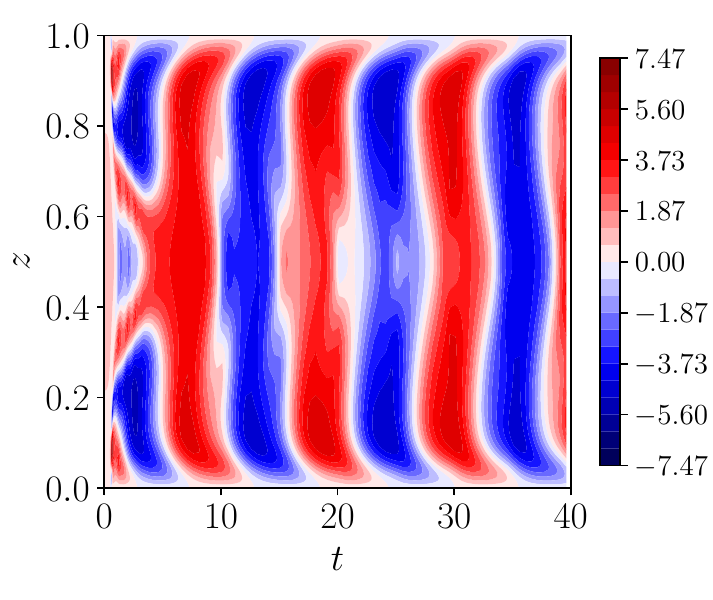}
		\caption{}
		\label{fig:contour_Ra80_s100}
	\end{subfigure}
	\par\bigskip 
	\begin{subfigure}{0.325\textwidth}
		\includegraphics[width=\linewidth]{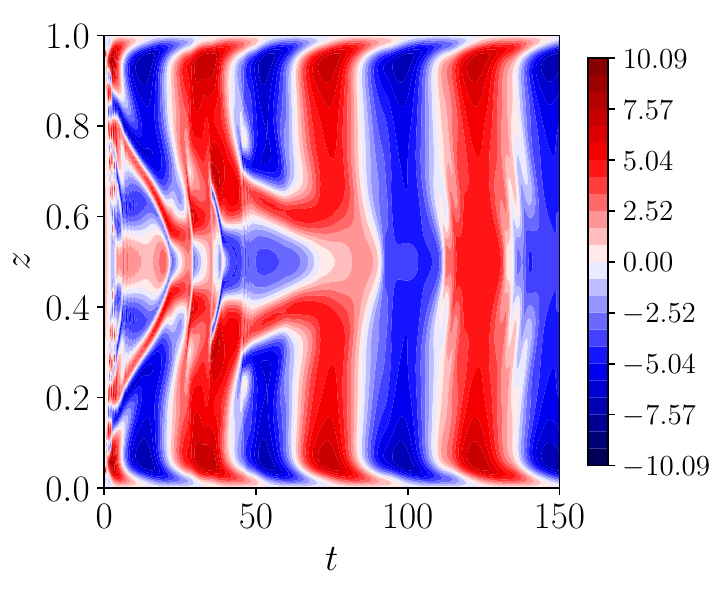}
		\caption{}
		\label{fig:contour_ekm6_ra150}
	\end{subfigure}
	\begin{subfigure}{0.325\textwidth}
		\includegraphics[width=\linewidth]{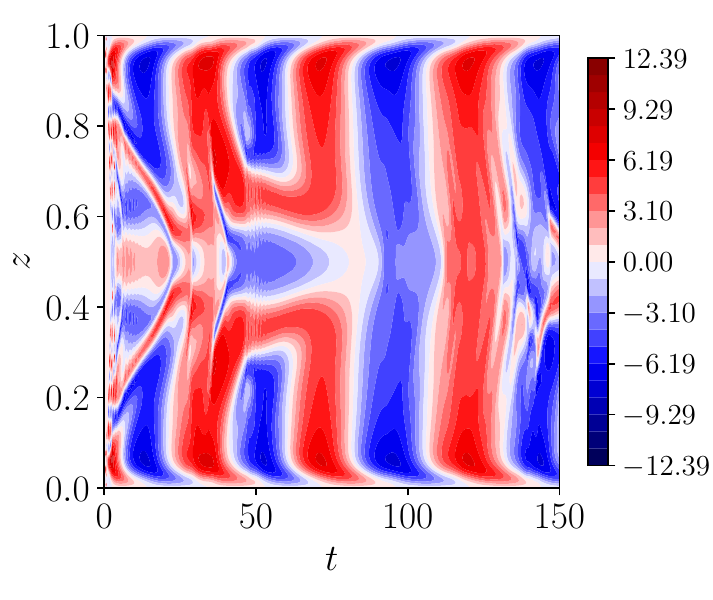}
		\caption{}
		\label{fig:contour_Ra150_s20}
	\end{subfigure}	
	\begin{subfigure}{0.325\textwidth}
		\includegraphics[width=\linewidth]{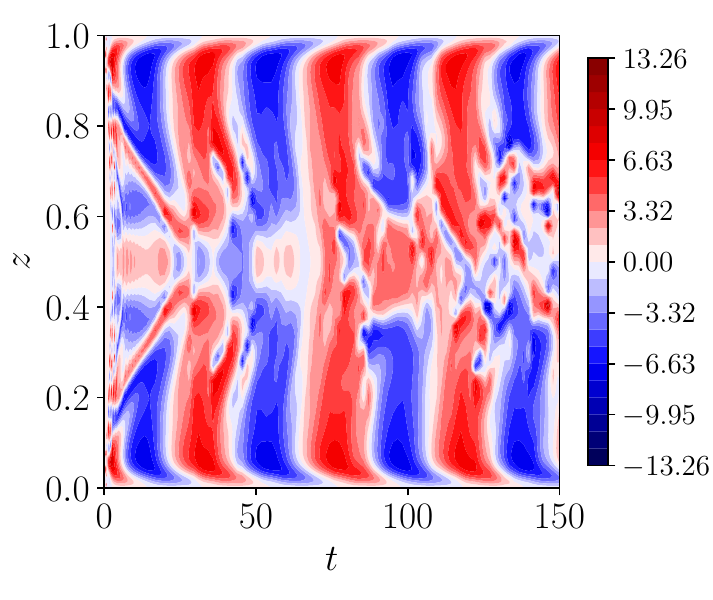}
		\caption{}
		\label{fig:contour_Ra150_s100}
	\end{subfigure}
	\caption{$B_x$ component profiles in space and time for $\revb{\widetilde{Ra}}=80, E=10^{-4}$, scaling factor $f=2.5$ on the top, $\revb{\widetilde{Ra}}=150, E=10^{-6}$, scaling factor $f=2.0$ at the bottom: a) reference solution, b) $s = 20$, c) $s = 100$, d) exact solution, e)  $s = 20$, f) $s = 100$.}
	\label{fig:contourBx}
\end{figure}

\begin{figure}[ht!]
	\begin{subfigure}{0.49\textwidth}
		\includegraphics[width=\linewidth]{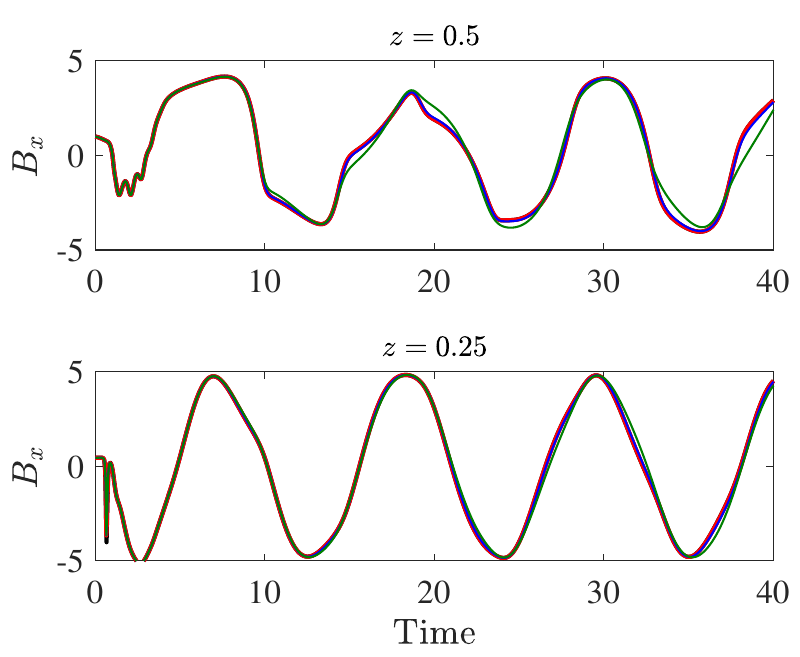}
		\caption{}
		\label{fig:ekm4_ra80_bx}
	\end{subfigure}
	\begin{subfigure}{0.49\textwidth}
		\includegraphics[width=\linewidth]{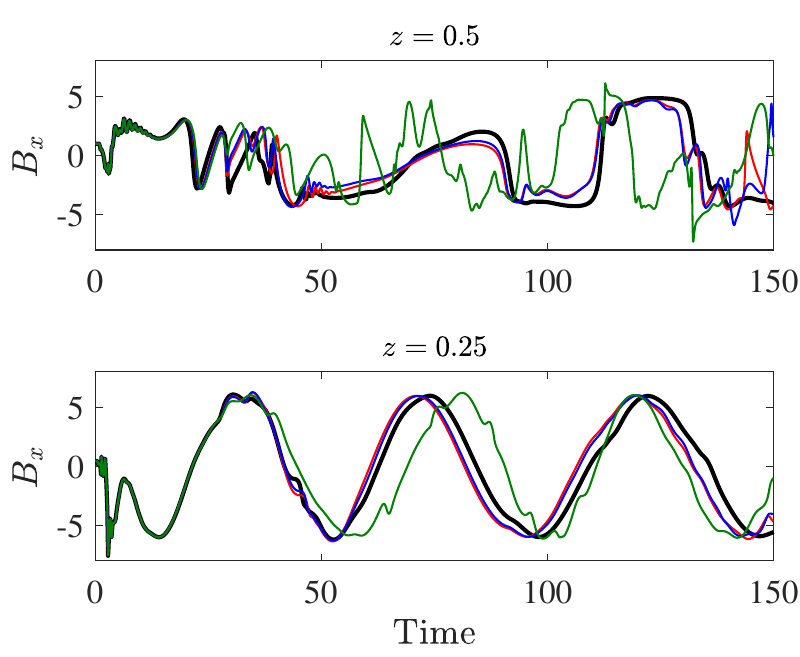}
		\caption{}
		\label{fig:ekm6_ra150_bx}
	\end{subfigure}
	\caption{\reva{Vertical slices of the $B_x$ component at $z = 0.25$ and $z = 0.5$ against time:  a) $\widetilde{Ra}=80, E=10^{-4}$, scaling factor $f = 2.5$, b) $\widetilde{Ra}=150, E=10^{-6}$, scaling factor $f = 2.0$. Reference solution (black), $s=20$ (red), $s=50$ (blue), $s=100$ (green).}}
	\label{fig:vert_slices_bx}
\end{figure}

Fig.~\ref{fig:contourBx} presents 2D contour plots of the mean magnetic field $ B_x $ component in space and time for different averaging windows and simulation parameters.
The reference solution for $\revb{\widetilde{Ra}} = 80, E=10^{-4}$ parameters is presented \reva{in} Fig.~\ref{fig:contour_ekm4_ra80}. Simulation results with averaging windows $ s = 20$, $s = 100 $ and $ f = 2.5 $ have the same periodic structure, however minor disruptions can be seen on the equatorial plane at $ z = 0.5 $ for approximate times $ t=20 $ and $ t = 30 $ \reva{in} Fig.~\ref{fig:contour_Ra80_s100}. Vertical slices for this case at $ z = 0.25 $ and 0.5 are presented \reva{in} Fig.~\ref{fig:ekm4_ra80_bx}. The reference solution for $\revb{\widetilde{Ra}}=150, E=10^{-6}$ presented \reva{in} Fig.~\ref{fig:contour_ekm6_ra150} and numerical simulations with averaging windows $ s = 20$, $s = 100 $ using $ f = 2.0 $ displayed \reva{in} Fig.~\ref{fig:contour_Ra150_s20} and Fig.~\ref{fig:contour_Ra150_s100}.

It is clearly seen that large number of time steps on the fast scale at high $ \revb{\widetilde{Ra}} $  produces inaccurate results with disrupted structure and small scale ripples on the mean magnetic field. Time evolution of vertical slices for these simulations are presented \reva{in} Fig.~\ref{fig:ekm6_ra150_bx}. The results at the equatorial plane using $ \Delta t = 5\cdot10^{-4} $ and $ s=100 $ show large deviations from the reference solution, yet the  outer regions for $ z = 0.25 $ show good agreement with the reference albeit with shifted phases. The discrepancy seen near $z = 0.5$ appears related to the behavior shown in Fig.~\ref{fig:contourBx}, as the disturbances propagate towards the middle of the vertical domain.

\begin{figure}[ht!]
	\begin{subfigure}{0.49\textwidth}
		\includegraphics[width=\linewidth]{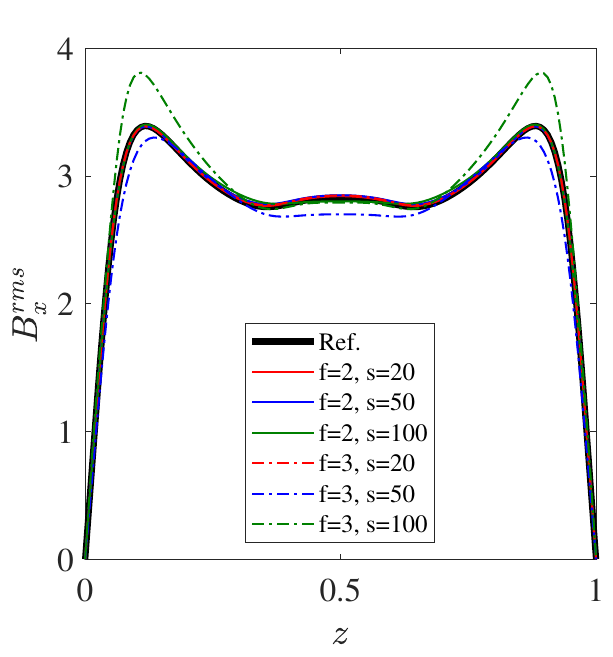}
	\end{subfigure}
	\begin{subfigure}{0.49\textwidth}
		\includegraphics[width=\linewidth]{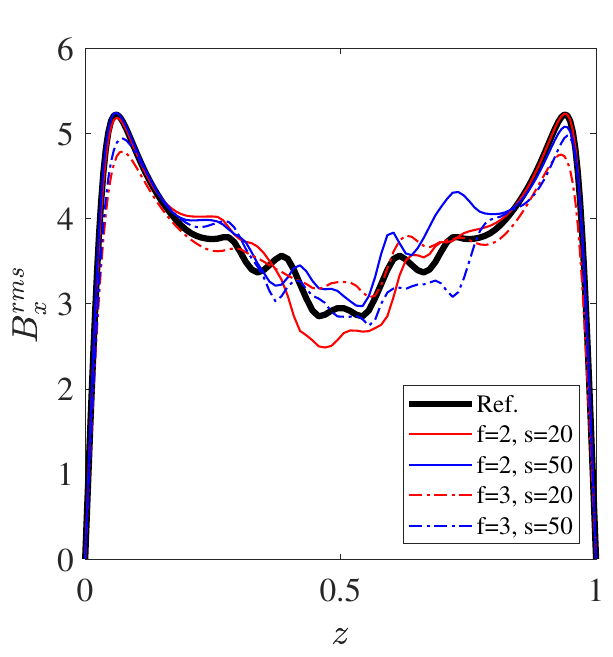}
	\end{subfigure}
	\caption{\reva{Time averaged vertical profiles of $B_x^{rms}$ for $\widetilde{Ra}=80, E=10^{-4}$ on the left and for $\widetilde{Ra}=150, E=10^{-6}$ on the right. Solid black line (Ref.) represents direct solution of the asymptotic model, $f$ is scaling factor and $s$ is number of fast steps.}}
	\label{fig:vert_bx}
\end{figure}

\begin{figure}[ht!]
	\begin{subfigure}{0.49\textwidth}
		\includegraphics[width=\linewidth]{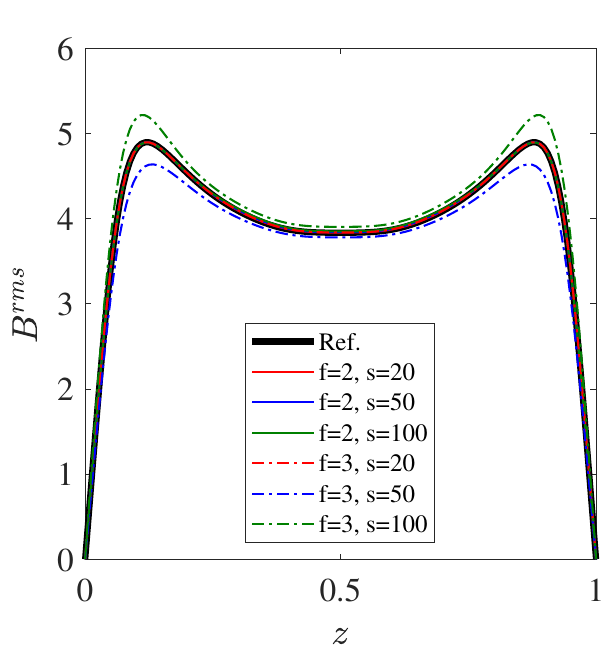}
	\end{subfigure}
	\begin{subfigure}{0.49\textwidth}
		\includegraphics[width=\linewidth]{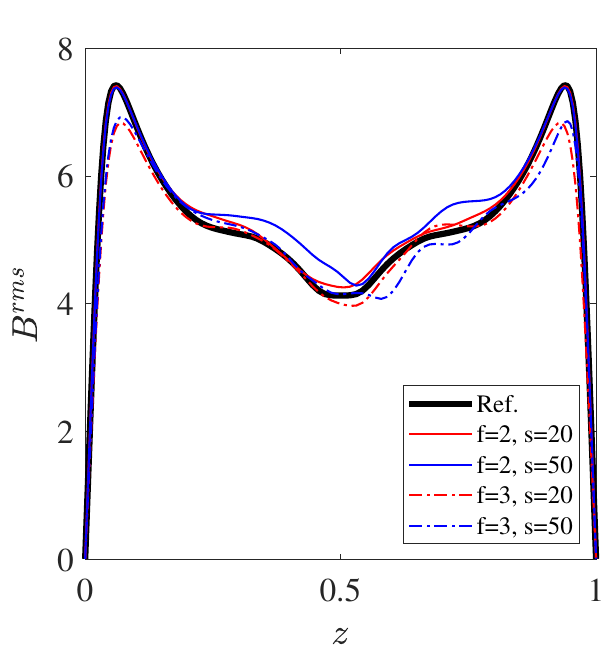}
	\end{subfigure}
	\caption{\reva{Time averaged vertical profiles of $B^{rms}$ for $\widetilde{Ra}=80, E=10^{-4}$ on the left and for $\widetilde{Ra}=150, E=10^{-6}$ on the right. Solid black line (Ref.) represents direct solution of the asymptotic model, $f$ is scaling factor and $s$ is number of fast steps.}}
	\label{fig:vert_b2}
\end{figure}

For additional analysis we've assessed time averaged vertical profiles of $ B_x $ and $ B = B_x^2 + B_y^2 $ for different values of $ \revb{\widetilde{Ra}}, E $ and scaling factors for the slow step which are presented \reva{in} Fig.~\ref{fig:vert_bx}-\ref{fig:vert_b2}. We compute the averaged values as
\begin{equation}
f^{rms} = \sqrt{\dfrac{1}{T} \int_{0}^{T} [f(t)]^2 dt}.
\end{equation}
The obtained vertical profiles for both low and high $ \revb{\widetilde{Ra}} $ are consistent with the time evolution graphs presented above for different sizes of the averaging windows and number of steps on the slow scale. The reduction strategy delivers highly accurate results at $ \revb{\widetilde{Ra}} = 80 $ using $ f = 2.0 $ and is capable of capturing the right dynamics for scaling factors up to $3.0$. At $ \revb{\widetilde{Ra}} = 150 $ the scheme fails to capture the fine structure of vertical profiles especially in at the midpoint for $ B_x^{rms} $ \reva{in} Fig.~\ref{fig:vert_bx} but this can be recovered by decreasing the micro step $ \Delta t $ in favour of greater computational work.

\subsection{Performance}
\begin{figure}[ht!]
    \centering
	\begin{subfigure}{0.45\textwidth}
		\includegraphics[width=\linewidth]{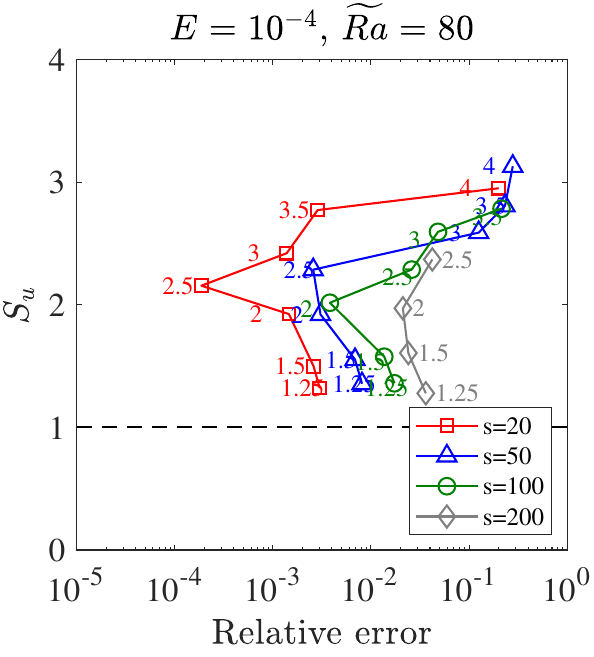}
		\caption{}
	\end{subfigure}
	\begin{subfigure}{0.45\textwidth}
		\includegraphics[width=\linewidth]{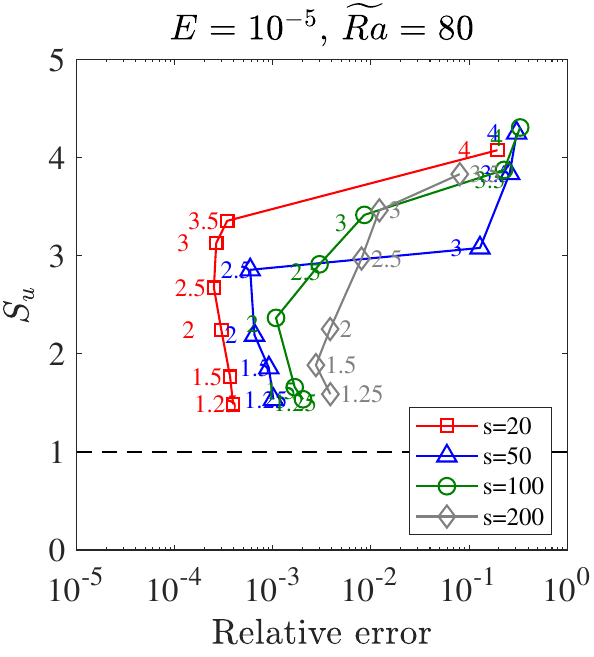}
		\caption{}
	\end{subfigure}
	\begin{subfigure}{0.45\textwidth}
		\includegraphics[width=\linewidth]{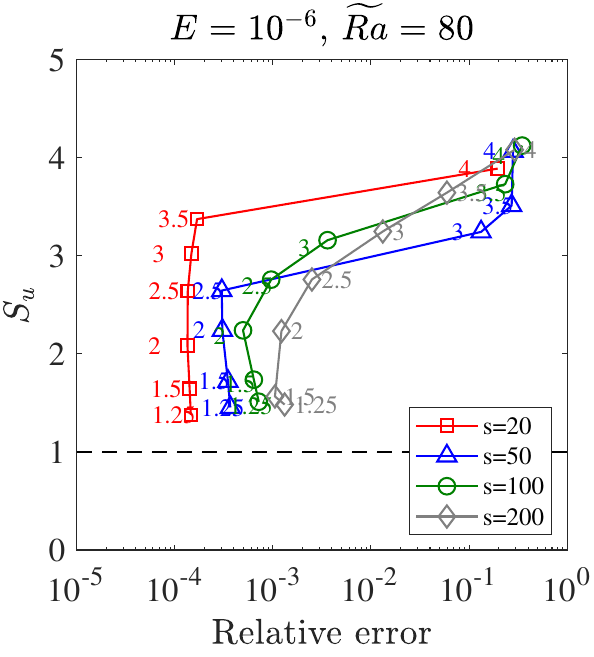}
		\caption{}
	\end{subfigure}
	\begin{subfigure}{0.45\textwidth}
		\includegraphics[width=\linewidth]{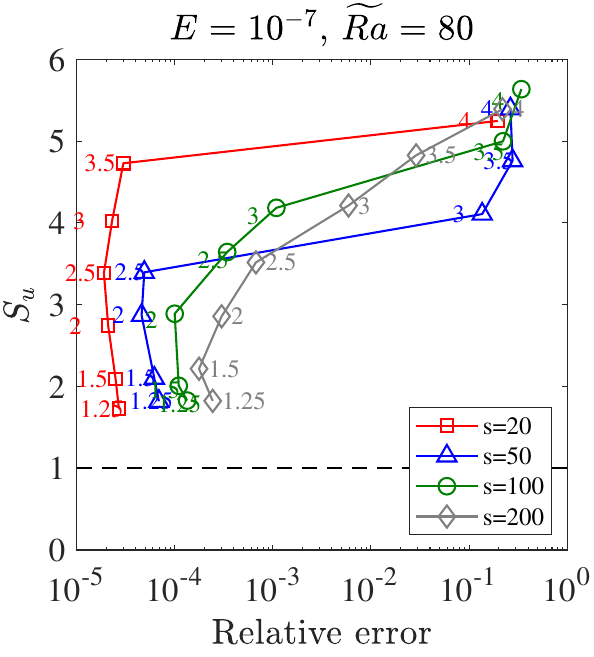}
		\caption{}
	\end{subfigure}
	\caption{\reva{Ratio $S_u$ of time needed to solve the system applying direct solution divided by time needed to solve the system using HMM-like approach. Simulations are carried out for different Ekman numbers and $ \widetilde{Ra}=80 $: a) $ E=10^{-4}$, b) $E=10^{-5}$, c) $E=10^{-6}$, d) $E=10^{-7}$.}}
	\label{fig:effort}
\end{figure}

To assess the computational effort we measure total runtime $ T_{sim} $ needed to solve the system until desired time $ t_{end} $ using different sizes of the slow step $\delta T$, scaling factor $f$ and averaging window $s$. Fig.~\ref{fig:effort} shows multiple curves obtained from simulations at different values of $ E $ using fixed micro step $ \Delta t = 5\cdot10^{-4} $ and $ \revb{\widetilde{Ra}} = 80 $. 
All four graphs present relative error of the magnetic energy against the $S_u$ value, where $S_u$ defines the ratio between time needed to obtain the direct solution of the asymptotic model (the reference solution) and time required to solve it by the HMM-like scheme. A value $S_u > 1$ indicates that HMM-scheme delivers computational gains and requires fewer evaluations to get the approximated solution. We compute the relative error using formula (\ref{E_rel}). Each graph \reva{in} Fig.~\ref{fig:effort} contains coloured lines associated with different averaging windows and markers defining the scaling factor $f$. 
\begin{figure}[ht!]
	\begin{subfigure}{0.49\textwidth}
	\includegraphics[width=\linewidth]{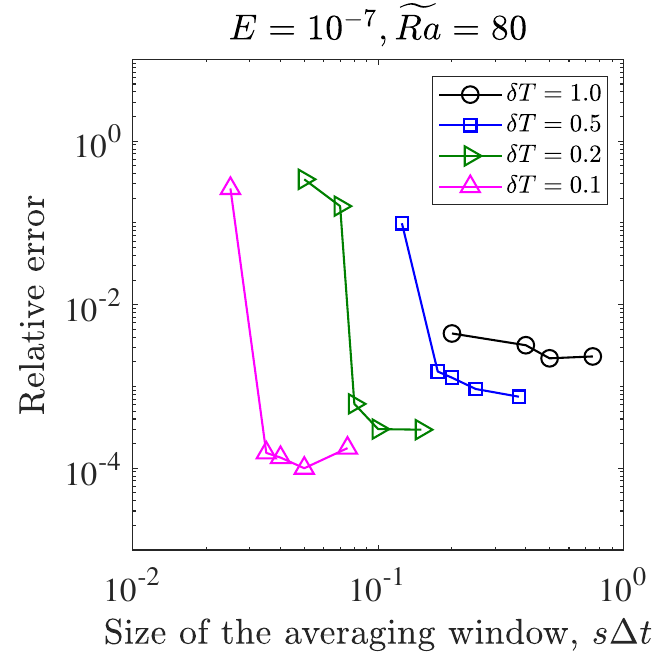}
	\caption{}
	\end{subfigure}
	\begin{subfigure}{0.49\textwidth}
	\includegraphics[width=\linewidth]{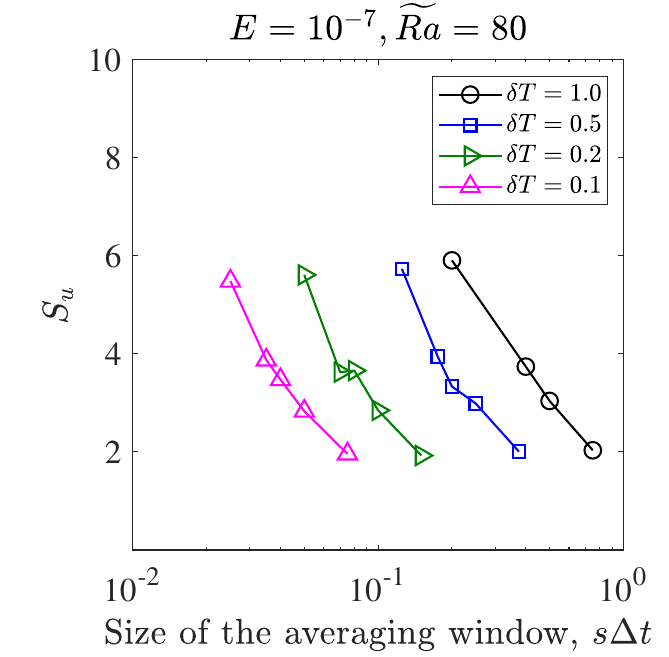}
	\caption{}
	\end{subfigure}
	\caption{\reva{Relative error of the magnetic energy (on the left) and ratio $S_u$ of time needed to solve the system applying direct solution divided by time needed to solve the system using HMM-like approach (on the right). Simulations are carried out for different fixed $\delta T$ while size of the averaging window $s \Delta t$ is changing.}}
	\label{fig:fixed_macro_step}
\end{figure}
As shown in Fig.~\ref{fig:effort} the HMM-scheme can deliver highly accurate solutions while requiring less computational work compared to direct simulation. The black curves represent the results for the smallest averaging window $ s = 20 $ and show stable results with scaling factors up to $ f = 3.5 $, indicating that it is possible to take a slow step $ 3.5 $ times larger than the averaging window. The increase of the averaging window results in a smaller number of the slow steps to be taken on certain time interval however this does not lead to substantial reduction of integration time at fixed $ f $. For the case with $ E = 10^{-4} $ and $ f = 2.0 $, for example, we have $ T_{sim}  = 1203$ for $ s = 20$ and $ T_{sim}  = 1191$ for $ s = 200$. The main reason is that the evaluation of equations on the fast scale is highly expensive in relation to the equations for the mean magnetic field. A decrease in the Ekman number positively affects the stability and accuracy of the method and allows for further reduction of the computational work as expected.

The optimal time stepping strategy to maximize both the $S_u$ value and accuracy is to use a small averaging window to capture the fast dynamics.
By increasing the number of fast steps it is possible to get an additional slight advantage in runtime but in favour of lower accuracy which is shown \reva{in} Fig.~\ref{fig:effort}. \reva{Alternative simulations are presented in Fig.~\ref{fig:fixed_macro_step} where we fixed macro step $\delta T$ and varied size of the averaging window. This allows for a reduction of the integration time by a factor of up to \reva{6 for $E=10^{-7}$}.}

\begin{figure}[ht!]
	\centering
	\begin{subfigure}{0.5\textwidth}
	\includegraphics[width=\linewidth]{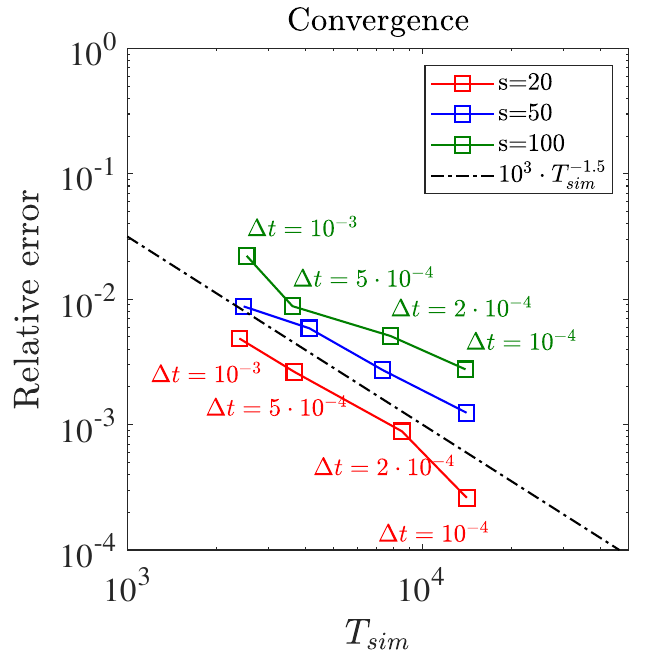}
	\end{subfigure}
	\caption{\reva{Convergence rate of the magnetic energy to reference data against runtime for different micro steps $ \Delta t$ and averaging windows $s \Delta t$ with fixed $ E = 10^{-6}, \widetilde{Ra}=130 $ and scaling factor $f=2$. Macro step changes automatically according to formula (\ref{slow_step}).}}
	\label{fig:convergence}
\end{figure}

In the next example we took slightly higher Rayleigh number ($ \revb{\widetilde{Ra}} = 130 $) in order to \reva{increase the effect of chaotic modulated solutions} and test how the change of micro step affects the accuracy and performance. Fig.~\ref{fig:convergence} shows multiple plots for relative error of the magnetic energy obtained from simulations with different micro steps using averaging windows of $ s = 20 $, $ s  = 50 $ \reva{and $s=100$ with fixed scaling factor $f=2$, i.e. $\delta T = 2 s \Delta t$}. 
\reva{Three} coloured lines show how the time stepping scheme converges when decreasing the micro step\reva{,} and the amount of computational work required. \reva{All curves have relatively the same slope which corresponds to a log–log plot with slope $m=-1.5$ (dash dot line).} \revb{The relative error scales as $T_{sim}^{-1.5}$ which means that reduction in the relative error by one order approximately corresponds to a factor 4 increase in the runtime. For example, the scheme with $s=20$ and $\Delta t = 5 \cdot 10^{-4}$ delivers $E_{rel} = 0.00263$ with $T_{sim}=$ \SI{3677}{\second} whereas more accurate simulation with $E_{rel} = 0.000261$ requires $T_{sim}=$ \SI{14119}{\second}. Smaller averaging windows (red curve) show better accuracy comparing to the green curve since the macro step increases with increasing $s$.}

\begin{figure}[ht!]
	\begin{subfigure}{0.49\textwidth}
	\includegraphics[width=\linewidth]{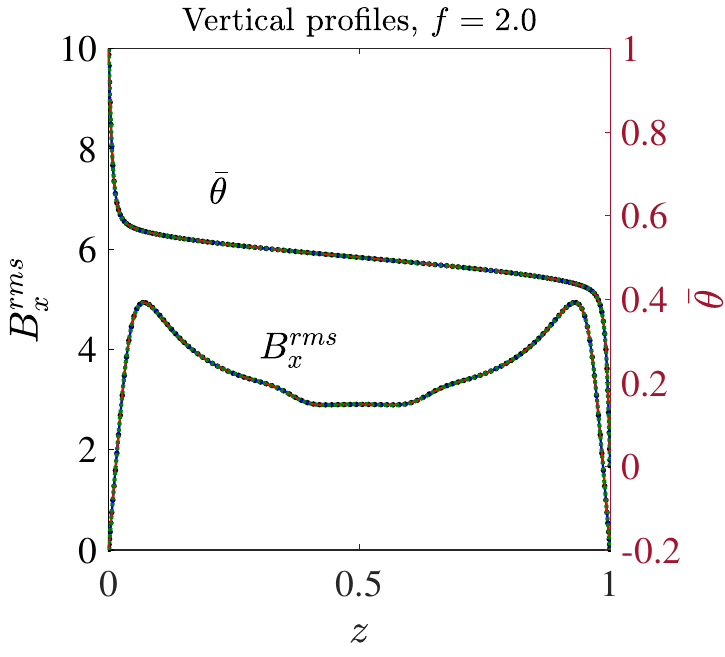}
	\caption{}
	\end{subfigure}
	\begin{subfigure}{0.49\textwidth}
	\includegraphics[width=\linewidth]{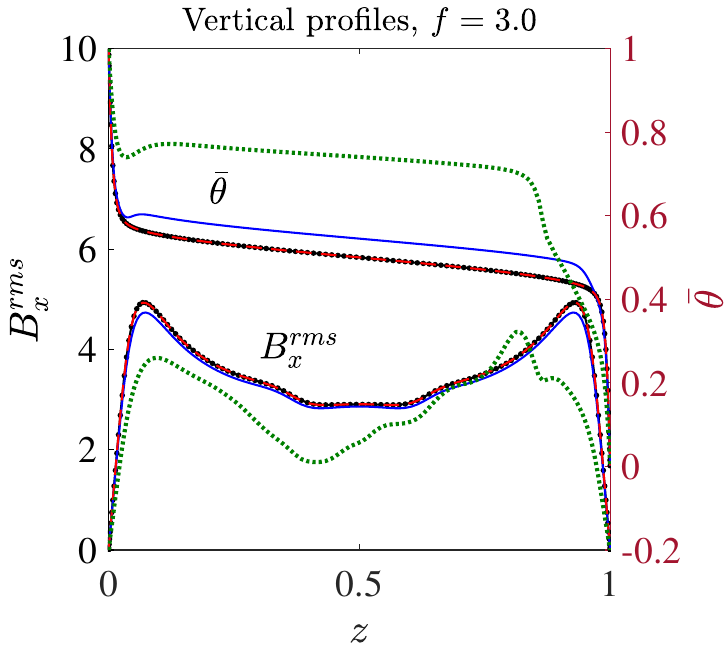}
	\caption{}
	\end{subfigure}
	\caption{Vertical profiles of $ B_x $ component (left Y axis) and mean temperature $ \bar{\theta} $ (right Y axis) for different micro steps $ \Delta t $ and scaling factors a) $f=2.0$ , b) $f=3.0$ with fixed $ s = 50, E = 10^{-6}, \widetilde{Ra}=130 $. Reference solution -- black dots, $\Delta t = 10^{-4}$ -- solid black, $\Delta t = 2 \cdot 10^{-4}$ -- solid blue, $\Delta t = 5 \cdot 10^{-4}$ -- dashed red, $\Delta t = 10^{-3}$ -- dotted green. }
	\label{fig:vert_profMult}
\end{figure}

To illustrate how the accuracy of the time dependent mean magnetic field energy corresponds to vertical profiles we refer to Fig.~\ref{fig:vert_profMult} which presents solutions of $ B_x $ component and mean temperature $ \bar{\theta} $ at fixed averaging window $ s = 50 $.
Left graph \reva{in} Fig.~\ref{fig:vert_profMult} contains two $y$-axes and shows the agreement of the  profiles with reference solution when using a scaling factor $f=2.0$ for the slow step. Results of numerical simulations with $ f = 3.0 $ show discrepancies in the profiles, where a maximum deviation is observed for largest micro step $ \Delta t = 10^{-3} $ (green line) and the relative error is the order of $ 10^{-1} $. Smaller values of $ \Delta t $ give better results in accuracy having \reva{$E_{rel} \sim 10^{-2}$} at  $ \Delta t = 5\cdot10^{-3} $ (red line) and \reva{$E_{rel} \sim 10^{-3}$} at $ \Delta t = 10^{-4} $ (black line).
\section{Conclusions}\label{conclusions}
\revc{In this paper we have applied a multi-scale integration method to the quasi-geostrophic self-consistent dynamo model. The presented multi-scale scheme is an example of the heterogeneous multiscale modelling method, which is a general methodology for the efficient numerical computations of problems that incorporate multiple temporal scales. Though we believe the method we present is generally applicable to nonlinear systems with a separation of timescales, it is particularly effective for quasilinear systems --- as described below.}

A large degree of scale separation in a system can lead to difficulties in simulation, due to a lack of knowledge of the interaction between scales and a large computational cost.  However, the use of techniques for handling mutliple scales in the models can provide the opportunity for large gains in computational efficiency. In particular, by combining a macrosolver employing a large integration time step on the slower dynamics and a microsolver for the full multi-dimensional system, the computational cost can be reduced while maintaining a large degree of accuracy of the solution.  

\revc{The system of equations for the QGDM describes a type of quasilinear system having fast dynamical modes that are coupled to the evolution of slow variables. Such a quasilinear system would undergo an instability in the absence of feedback on the slow variables.} The gains in efficiency for simulating the QGDM provide insight into the potential for HMM methods in solving multiscale PDEs and highlight some limitations for accuracy. The key to the efficiency of the HMM-like method used here is to separate the timestepping and solving of the slow, mean magnetic scale computation from the fast, convective scale computation.  The fast equations are calculated on the smaller timestep \reva{$\Delta t$} whereas the slower dynamics are updated only every \reva{$\delta T $} steps, and the fluxes are integrated using a smoothing weight function over the period of time the slow dynamics are held steady. The difference between \reva{$\delta T $} and \reva{$\Delta t$} provides for a computational savings, as the larger \reva{$\delta T $} is, the fewer slow scale solves are necessary. 

The direct computation of the QGDM compared with the results from applying this HMM-like method indicates that this method accurately captures the dynamics within varying tolerances.  For sufficiently small \reva{$\delta T $}, both the instantaneous and  averaged dynamics show strong agreement, including the behavior of the Nusselt number, energy and vertical profiles.  For larger $f > 2.5$ (and therefore larger \reva{$\delta T $}), the agreement is reduced for the instantaneous results, however the relative error is still bounded below roughly $10^{-1}$.

The derivation of the QGDM relies on the model applying in the limit of rapid rotation and $E \ll 1$, a regime in which the scale separation is most pronounced between the fast and slow dynamics (recalling that the scale separation is $\mathcal{O}(E^{1/2})$).  Thus, it is expected that this HMM method will be able to provide larger gains in efficiency for lower $E$ as more separation requires fewer solves.  Results at the lowest $E = 10^{-7}$ show the largest efficiency gains for the system, with a speedup of roughly \reva{six} times and a relative error $< 10^{-2}$.  Extrapolating from this result implies that for even lower $E$, the gains in efficiency will improve.


\bibliographystyle{elsarticle-num-names}
\bibliography{refs.bib}

\begin{thebibliography}{35}
\expandafter\ifx\csname natexlab\endcsname\relax\def\natexlab#1{#1}\fi
\providecommand{\url}[1]{\texttt{#1}}
\providecommand{\href}[2]{#2}
\providecommand{\path}[1]{#1}
\providecommand{\DOIprefix}{doi:}
\providecommand{\ArXivprefix}{arXiv:}
\providecommand{\URLprefix}{URL: }
\providecommand{\Pubmedprefix}{pmid:}
\providecommand{\doi}[1]{\href{http://dx.doi.org/#1}{\path{#1}}}
\providecommand{\Pubmed}[1]{\href{pmid:#1}{\path{#1}}}
\providecommand{\bibinfo}[2]{#2}
\ifx\xfnm\relax \def\xfnm[#1]{\unskip,\space#1}\fi
\bibitem[{Brandt(2002)}]{Multiscale_modelling_Brandt}
\bibinfo{author}{A.~Brandt},
\newblock \bibinfo{title}{Multiscale scientific computation: Review 2001},
\newblock in: \bibinfo{editor}{T.~J. Barth}, \bibinfo{editor}{T.~Chan},
  \bibinfo{editor}{R.~Haimes} (Eds.), \bibinfo{booktitle}{Multiscale and
  Multiresolution Methods}, \bibinfo{publisher}{Springer Berlin Heidelberg},
  \bibinfo{address}{Berlin, Heidelberg}, \bibinfo{year}{2002}, pp.
  \bibinfo{pages}{3--95}.
\bibitem[{{E}(2011)}]{WE_2011}
\bibinfo{author}{W.~{E}}, \bibinfo{title}{{Principles of Multiscale Modeling}},
  \bibinfo{publisher}{Cambridge University Press}, \bibinfo{year}{2011}.
\bibitem[{Chen and Doolen(1998)}]{HMM_fluids}
\bibinfo{author}{S.~Chen}, \bibinfo{author}{G.~D. Doolen},
\newblock \bibinfo{title}{Lattice boltzmann method for fluid flows},
\newblock \bibinfo{journal}{Annual Review of Fluid Mechanics}
  \bibinfo{volume}{30} (\bibinfo{year}{1998}) \bibinfo{pages}{329--364}.
  \DOIprefix\doi{10.1146/annurev.fluid.30.1.329}.
\bibitem[{Malecha et~al.(2014)Malecha, Chini, and Julien}]{HMM_fluids_Malecha}
\bibinfo{author}{Z.~Malecha}, \bibinfo{author}{G.~Chini},
  \bibinfo{author}{K.~Julien},
\newblock \bibinfo{title}{A multiscale algorithm for simulating
  spatially-extended langmuir circulation dynamics},
\newblock \bibinfo{journal}{Journal of Computational Physics}
  \bibinfo{volume}{271} (\bibinfo{year}{2014}) \bibinfo{pages}{131--150}.
  \DOIprefix\doi{10.1016/j.jcp.2013.07.003}.
\bibitem[{E and Li(2004)}]{HMM_gas_dyn}
\bibinfo{author}{W.~E}, \bibinfo{author}{X.~Li},
\newblock \bibinfo{title}{Analysis of the heterogeneous multiscale method for
  gas dynamics},
\newblock \bibinfo{journal}{Methods Appl. Anal.} \bibinfo{volume}{11}
  (\bibinfo{year}{2004}) \bibinfo{pages}{557--572}.
\bibitem[{Li and E(2005)}]{HMM_solids}
\bibinfo{author}{X.~Li}, \bibinfo{author}{W.~E},
\newblock \bibinfo{title}{Multiscale modeling of the dynamics of solids at
  finite temperature},
\newblock \bibinfo{journal}{Journal of the Mechanics and Physics of Solids}
  \bibinfo{volume}{53} (\bibinfo{year}{2005}) \bibinfo{pages}{1650--1685}.
  \DOIprefix\doi{https://doi.org/10.1016/j.jmps.2005.01.008}.
\bibitem[{Ulz(2014)}]{HMM_molecular}
\bibinfo{author}{M.~H. Ulz},
\newblock \bibinfo{title}{The heterogeneous multiscale method as a framework
  for coupling the finite element method and molecular dynamics},
\newblock \bibinfo{journal}{PAMM} \bibinfo{volume}{14} (\bibinfo{year}{2014})
  \bibinfo{pages}{571--572}. \DOIprefix\doi{10.1002/pamm.201410273}.
\bibitem[{E and Engquist(2003)}]{HMM_orig_2003}
\bibinfo{author}{W.~E}, \bibinfo{author}{B.~Engquist},
\newblock \bibinfo{title}{The heterognous multiscale methods},
\newblock \bibinfo{journal}{Commun. Math. Sci.} \bibinfo{volume}{1}
  (\bibinfo{year}{2003}) \bibinfo{pages}{87--132}.
\bibitem[{Engquist and Tsai(2005)}]{HMM_orig_2005}
\bibinfo{author}{B.~Engquist}, \bibinfo{author}{Y.-H. Tsai},
\newblock \bibinfo{title}{Heterogeneous multiscale methods for stiff ordinary
  differential equations},
\newblock \bibinfo{journal}{Mathematics of Computation} \bibinfo{volume}{74}
  (\bibinfo{year}{2005}) \bibinfo{pages}{1707--1742}.
  \DOIprefix\doi{10.1090/s0025-5718-05-01745-x}.
\bibitem[{E et~al.(2007)E, Engquist, Li, Ren, and
  Vanden-Eijnden}]{HMM_review_2007}
\bibinfo{author}{W.~E}, \bibinfo{author}{B.~Engquist}, \bibinfo{author}{X.~Li},
  \bibinfo{author}{W.~Ren}, \bibinfo{author}{E.~Vanden-Eijnden},
\newblock \bibinfo{title}{Heterogeneous multiscale methods: A review},
\newblock \bibinfo{journal}{Communications in Computational Physics}
  \bibinfo{volume}{2} (\bibinfo{year}{2007}) \bibinfo{pages}{367--450}.
\bibitem[{Kevrekidis et~al.(2003)Kevrekidis, Gear, Hyman, Kevrekidid, Runborg,
  and Theodoropoulos}]{EqFree_kevrekidis_2003}
\bibinfo{author}{I.~G. Kevrekidis}, \bibinfo{author}{C.~W. Gear},
  \bibinfo{author}{J.~M. Hyman}, \bibinfo{author}{P.~G. Kevrekidid},
  \bibinfo{author}{O.~Runborg}, \bibinfo{author}{C.~Theodoropoulos},
\newblock \bibinfo{title}{Equation-free, coarse-grained multiscale computation:
  Enabling microscopic simulators to perform system-level analysis},
\newblock \bibinfo{journal}{Commun. Math. Sci.} \bibinfo{volume}{1}
  (\bibinfo{year}{2003}) \bibinfo{pages}{715--762}.
\bibitem[{Kevrekidis and Samaey(2009)}]{EqFree_2009}
\bibinfo{author}{I.~G. Kevrekidis}, \bibinfo{author}{G.~Samaey},
\newblock \bibinfo{title}{Equation-free multiscale computation: Algorithms and
  applications},
\newblock \bibinfo{journal}{Annual Review of Physical Chemistry}
  \bibinfo{volume}{60} (\bibinfo{year}{2009}) \bibinfo{pages}{321--344}.
  \DOIprefix\doi{10.1146/annurev.physchem.59.032607.093610}.
\bibitem[{Lee and Engquist(2014)}]{VSS_HMM}
\bibinfo{author}{Y.~Lee}, \bibinfo{author}{B.~Engquist},
\newblock \bibinfo{title}{Variable step size multiscale methods for stiff and
  highly oscillatory dynamical systems},
\newblock \bibinfo{journal}{Discrete \& Continuous Dynamical Systems - A}
  \bibinfo{volume}{34} (\bibinfo{year}{2014}) \bibinfo{pages}{1079--1097}.
  \DOIprefix\doi{10.3934/dcds.2014.34.1079}.
\bibitem[{Lee and Engquist(2015)}]{lee2015fast}
\bibinfo{author}{Y.~Lee}, \bibinfo{author}{B.~Engquist}, \bibinfo{title}{Fast
  integrators for dynamical systems with several temporal scales},
  \bibinfo{year}{2015}. \href{http://arxiv.org/abs/1510.05728}{{\tt
  arXiv:1510.05728}}.
\bibitem[{Tao et~al.(2010)Tao, Owhadi, and Marsden}]{Tao2010}
\bibinfo{author}{M.~Tao}, \bibinfo{author}{H.~Owhadi}, \bibinfo{author}{J.~E.
  Marsden},
\newblock \bibinfo{title}{Nonintrusive and structure preserving multiscale
  integration of stiff {ODEs}, {SDEs}, and hamiltonian systems with hidden slow
  dynamics via flow averaging},
\newblock \bibinfo{journal}{Multiscale Modeling {\&} Simulation}
  \bibinfo{volume}{8} (\bibinfo{year}{2010}) \bibinfo{pages}{1269--1324}.
  \DOIprefix\doi{10.1137/090771648}.
\bibitem[{Fatkullin and Vanden-Eijnden(2004)}]{FATKULLIN2004605}
\bibinfo{author}{I.~Fatkullin}, \bibinfo{author}{E.~Vanden-Eijnden},
\newblock \bibinfo{title}{A computational strategy for multiscale systems with
  applications to {Lorenz} 96 model},
\newblock \bibinfo{journal}{Journal of Computational Physics}
  \bibinfo{volume}{200} (\bibinfo{year}{2004}) \bibinfo{pages}{605--638}.
  \DOIprefix\doi{https://doi.org/10.1016/j.jcp.2004.04.013}.
\bibitem[{E et~al.(2009)E, Ren, and Vanden-Eijnden}]{E2009}
\bibinfo{author}{W.~E}, \bibinfo{author}{W.~Ren},
  \bibinfo{author}{E.~Vanden-Eijnden},
\newblock \bibinfo{title}{A general strategy for designing seamless multiscale
  methods},
\newblock \bibinfo{journal}{Journal of Computational Physics}
  \bibinfo{volume}{228} (\bibinfo{year}{2009}) \bibinfo{pages}{5437--5453}.
  \DOIprefix\doi{10.1016/j.jcp.2009.04.030}.
\bibitem[{Gear and Kevrekidis(2003)}]{PI_origin}
\bibinfo{author}{C.~W. Gear}, \bibinfo{author}{I.~G. Kevrekidis},
\newblock \bibinfo{title}{Projective methods for stiff differential equations:
  Problems with gaps in their eigenvalue spectrum},
\newblock \bibinfo{journal}{SIAM Journal on Scientific Computing}
  \bibinfo{volume}{24} (\bibinfo{year}{2003}) \bibinfo{pages}{1091--1106}.
  \DOIprefix\doi{10.1137/S1064827501388157}.
\bibitem[{Lee and Gear(2007)}]{PI_2nd_order}
\bibinfo{author}{S.~L. Lee}, \bibinfo{author}{C.~Gear},
\newblock \bibinfo{title}{Second-order accurate projective integrators for
  multiscale problems},
\newblock \bibinfo{journal}{Journal of Computational and Applied Mathematics}
  \bibinfo{volume}{201} (\bibinfo{year}{2007}) \bibinfo{pages}{258--274}.
  \DOIprefix\doi{10.1016/j.cam.2006.02.018}.
\bibitem[{Lafitte et~al.(2017)Lafitte, Melis, and Samaey}]{PI_high_order}
\bibinfo{author}{P.~Lafitte}, \bibinfo{author}{W.~Melis},
  \bibinfo{author}{G.~Samaey},
\newblock \bibinfo{title}{A high-order relaxation method with projective
  integration for solving nonlinear systems of hyperbolic conservation laws},
\newblock \bibinfo{journal}{Journal of Computational Physics}
  \bibinfo{volume}{340} (\bibinfo{year}{2017}) \bibinfo{pages}{1--25}.
  \DOIprefix\doi{10.1016/j.jcp.2017.03.027}.
\bibitem[{Vanden-Eijnden(2007)}]{Vanden-Eijnden2007}
\bibinfo{author}{E.~Vanden-Eijnden},
\newblock \bibinfo{title}{On {HMM}-like integrators and projective integration
  methods for systems with multiple time scales},
\newblock \bibinfo{journal}{Communications in Mathematical Sciences}
  \bibinfo{volume}{5} (\bibinfo{year}{2007}) \bibinfo{pages}{495–505}.
  \DOIprefix\doi{https://dx.doi.org/10.4310/CMS.2007.v5.n2.a14}.
\bibitem[{Tobias(2021)}]{tobias2019}
\bibinfo{author}{S.~Tobias},
\newblock \bibinfo{title}{The turbulent dynamo},
\newblock \bibinfo{journal}{Journal of Fluid Mechanics} \bibinfo{volume}{912}
  (\bibinfo{year}{2021}) \bibinfo{pages}{P1}.
  \DOIprefix\doi{10.1017/jfm.2020.1055}.
\bibitem[{Roberts and King(2013)}]{Roberts2013}
\bibinfo{author}{P.~H. Roberts}, \bibinfo{author}{E.~M. King},
\newblock \bibinfo{title}{On the genesis of the earth{\textquotesingle}s
  magnetism},
\newblock \bibinfo{journal}{Reports on Progress in Physics}
  \bibinfo{volume}{76} (\bibinfo{year}{2013}) \bibinfo{pages}{096801}.
  \DOIprefix\doi{10.1088/0034-4885/76/9/096801}.
\bibitem[{Glatzmaiers and Roberts(1995)}]{Glatzmaiers2_1995}
\bibinfo{author}{G.~A. Glatzmaiers}, \bibinfo{author}{P.~H. Roberts},
\newblock \bibinfo{title}{A three-dimensional self-consistent computer
  simulation of a geomagnetic field reversal},
\newblock \bibinfo{journal}{Nature} \bibinfo{volume}{377}
  (\bibinfo{year}{1995}) \bibinfo{pages}{203–209}.
  \DOIprefix\doi{https://doi.org/10.1038/377203a0}.
\bibitem[{{Schaeffer} et~al.(2017){Schaeffer}, {Jault}, {Nataf}, and
  {Fournier}}]{sjnf2017}
\bibinfo{author}{N.~{Schaeffer}}, \bibinfo{author}{D.~{Jault}},
  \bibinfo{author}{H.~C. {Nataf}}, \bibinfo{author}{A.~{Fournier}},
\newblock \bibinfo{title}{{Turbulent geodynamo simulations: a leap towards
  Earth's core}},
\newblock \bibinfo{journal}{Geophysical Journal International}
  \bibinfo{volume}{211} (\bibinfo{year}{2017}) \bibinfo{pages}{1--29}.
  \DOIprefix\doi{10.1093/gji/ggx265}.
\bibitem[{{Aubert} et~al.(2017){Aubert}, {Gastine}, and {Fournier}}]{agf2017}
\bibinfo{author}{J.~{Aubert}}, \bibinfo{author}{T.~{Gastine}},
  \bibinfo{author}{A.~{Fournier}},
\newblock \bibinfo{title}{{Spherical convective dynamos in the rapidly rotating
  asymptotic regime}},
\newblock \bibinfo{journal}{Journal of Fluid Mechanics} \bibinfo{volume}{813}
  (\bibinfo{year}{2017}) \bibinfo{pages}{558--593}.
  \DOIprefix\doi{10.1017/jfm.2016.789}.
\bibitem[{Calkins et~al.(2015)Calkins, Julien, Tobias, and
  Aurnou}]{Calkins2015}
\bibinfo{author}{M.~A. Calkins}, \bibinfo{author}{K.~Julien},
  \bibinfo{author}{S.~M. Tobias}, \bibinfo{author}{J.~M. Aurnou},
\newblock \bibinfo{title}{A multiscale dynamo model driven by quasi-geostrophic
  convection},
\newblock \bibinfo{journal}{Journal of Fluid Mechanics} \bibinfo{volume}{780}
  (\bibinfo{year}{2015}) \bibinfo{pages}{143--166}.
  \DOIprefix\doi{10.1017/jfm.2015.464}.
\bibitem[{Calkins et~al.(2016{\natexlab{a}})Calkins, Julien, Tobias, Aurnou,
  and Marti}]{Calkins2016a}
\bibinfo{author}{M.~A. Calkins}, \bibinfo{author}{K.~Julien},
  \bibinfo{author}{S.~M. Tobias}, \bibinfo{author}{J.~M. Aurnou},
  \bibinfo{author}{P.~Marti},
\newblock \bibinfo{title}{Convection-driven kinematic dynamos at low rossby and
  magnetic prandtl numbers: Single mode solutions},
\newblock \bibinfo{journal}{Phys. Rev. E} \bibinfo{volume}{93}
  (\bibinfo{year}{2016}{\natexlab{a}}) \bibinfo{pages}{023115}.
  \DOIprefix\doi{10.1103/PhysRevE.93.023115}.
\bibitem[{Calkins et~al.(2016{\natexlab{b}})Calkins, Long, Nieves, Julien, and
  Tobias}]{Calkins2016b}
\bibinfo{author}{M.~A. Calkins}, \bibinfo{author}{L.~Long},
  \bibinfo{author}{D.~Nieves}, \bibinfo{author}{K.~Julien},
  \bibinfo{author}{S.~M. Tobias},
\newblock \bibinfo{title}{Convection-driven kinematic dynamos at low rossby and
  magnetic prandtl numbers},
\newblock \bibinfo{journal}{Phys. Rev. Fluids} \bibinfo{volume}{1}
  (\bibinfo{year}{2016}{\natexlab{b}}) \bibinfo{pages}{083701}.
  \DOIprefix\doi{10.1103/PhysRevFluids.1.083701}.
\bibitem[{Plumley et~al.(2018)Plumley, Calkins, Julien, and
  Tobias}]{plumley2018}
\bibinfo{author}{M.~Plumley}, \bibinfo{author}{M.~A. Calkins},
  \bibinfo{author}{K.~Julien}, \bibinfo{author}{S.~M. Tobias},
\newblock \bibinfo{title}{Self-consistent single mode investigations of the
  quasi-geostrophic convection-driven dynamo model},
\newblock \bibinfo{journal}{Journal of Plasma Physics} \bibinfo{volume}{84}
  (\bibinfo{year}{2018}) \bibinfo{pages}{735840406}.
  \DOIprefix\doi{10.1017/S0022377818000831}.
\bibitem[{Michel and Chini(2019)}]{Michel2019}
\bibinfo{author}{G.~Michel}, \bibinfo{author}{G.~P. Chini},
\newblock \bibinfo{title}{Multiple scales analysis of slow{\textendash}fast
  quasi-linear systems},
\newblock \bibinfo{journal}{Proceedings of the Royal Society A: Mathematical,
  Physical and Engineering Sciences} \bibinfo{volume}{475}
  (\bibinfo{year}{2019}) \bibinfo{pages}{20180630}. \URLprefix
  \url{https://doi.org/10.1098/rspa.2018.0630}.
  \DOIprefix\doi{10.1098/rspa.2018.0630}.
\bibitem[{Burns et~al.(2020)Burns, Vasil, Oishi, Lecoanet, and Brown}]{dedalus}
\bibinfo{author}{K.~J. Burns}, \bibinfo{author}{G.~M. Vasil},
  \bibinfo{author}{J.~S. Oishi}, \bibinfo{author}{D.~Lecoanet},
  \bibinfo{author}{B.~P. Brown},
\newblock \bibinfo{title}{Dedalus: A flexible framework for numerical
  simulations with spectral methods},
\newblock \bibinfo{journal}{Phys. Rev. Research} \bibinfo{volume}{2}
  (\bibinfo{year}{2020}) \bibinfo{pages}{023068}.
  \DOIprefix\doi{10.1103/PhysRevResearch.2.023068}.
\bibitem[{Ascher et~al.(1997)Ascher, Ruuth, and Spiteric}]{rk443}
\bibinfo{author}{U.~M. Ascher}, \bibinfo{author}{S.~J. Ruuth},
  \bibinfo{author}{R.~J. Spiteric},
\newblock \bibinfo{title}{Implicit-explicit runge-kutta methods for
  time-dependent partial differential equations},
\newblock \bibinfo{journal}{Applied Numerical Mathematics} \bibinfo{volume}{25}
  (\bibinfo{year}{1997}) \bibinfo{pages}{151--167}.
  \DOIprefix\doi{10.1016/S0168-9274(97)00056-1}.
\bibitem[{Wang and Ruuth(2008)}]{sbdf2}
\bibinfo{author}{D.~Wang}, \bibinfo{author}{S.~J. Ruuth},
\newblock \bibinfo{title}{Variable step-size implicit-explicit linear multistep
  methods for time-dependent partial differential equations},
\newblock \bibinfo{journal}{Journal of Computational Mathematics}
  \bibinfo{volume}{26} (\bibinfo{year}{2008}) \bibinfo{pages}{838--855}.
\bibitem[{Chandrasekhar(1961)}]{chandrasekhar1961}
\bibinfo{author}{S.~Chandrasekhar}, \bibinfo{title}{Hydrodynamic and
  Hydromagnetic Stability}, \bibinfo{publisher}{Oxford University Press},
  \bibinfo{year}{1961}.

\end{thebibliography}

\end{document}